\documentstyle[psfig]{mn}

\newif\ifAMStwofonts

\ifoldfss

  \ifCUPmtlplainloaded \else

    \NewTextAlphabet{textbfit} {cmbxti10} {}

    \NewTextAlphabet{textbfss} {cmssbx10} {}

    \NewMathAlphabet{mathbfit} {cmbxti10} {} 

    \NewMathAlphabet{mathbfss} {cmssbx10} {} 

  \fi

  \ifAMStwofonts

    \ifCUPmtlplainloaded \else

      \NewSymbolFont{upmath} {eurm10}

      \NewSymbolFont{AMSa} {msam10}

      \NewMathSymbol{\upi}     {0}{upmath}{19}

      \NewMathSymbol{\umu}     {0}{upmath}{16}

      \NewMathSymbol{\upartial}{0}{upmath}{40}

      \NewMathSymbol{\leqslant}{3}{AMSa}{36}

      \NewMathSymbol{\geqslant}{3}{AMSa}{3E}

      \let\leq=\leqslant \let\le=\leqslant

       \let\ge=\geqslant

    \fi

  \fi

\fi 

\ifnfssone

  \newmathalphabet{\mathit}

  \addtoversion{normal}{\mathit}{cmr}{m}{it}

  \addtoversion{bold}{\mathit}{cmr}{bx}{it}

  \newmathalphabet{\mathbfit} 

  \addtoversion{normal}{\mathbfit}{cmr}{bx}{it}

  \addtoversion{bold}{\mathbfit}{cmr}{bx}{it}

  \newmathalphabet{\mathbfss} 

  \addtoversion{normal}{\mathbfss}{cmss}{bx}{n}

  \addtoversion{bold}{\mathbfss}{cmss}{bx}{n}

  \ifAMStwofonts

    \ifCUPmtlplainloaded \else

      \UseAMStwoboldmath

      \makeatletter

      \new@mathgroup\upmath@group

      \define@mathgroup\mv@normal\upmath@group{eur}{m}{n}

      \define@mathgroup\mv@bold\upmath@group{eur}{b}{n}

      \edef\UPM{\hexnumber\upmath@group}

      \new@mathgroup\amsa@group

      \define@mathgroup\mv@normal\amsa@group{msa}{m}{n}

      \define@mathgroup\mv@bold\amsa@group{msa}{m}{n}

      \edef\AMSa{\hexnumber\amsa@group}

      \makeatother

      \mathchardef\upi="0\UPM19

      \mathchardef\umu="0\UPM16

      \mathchardef\upartial="0\UPM40

      \mathchardef\leqslant="3\AMSa36

      \mathchardef\geqslant="3\AMSa3E

      \let\leq=\leqslant \let\le=\leqslant

       \let\ge=\geqslant

    \fi

  \fi

\fi 

\ifnfsstwo

  \DeclareMathAlphabet{\mathbfit}{OT1}{cmr}{bx}{it}

  \SetMathAlphabet\mathbfit{bold}{OT1}{cmr}{bx}{it}

  \DeclareMathAlphabet{\mathbfss}{OT1}{cmss}{bx}{n}

  \SetMathAlphabet\mathbfss{bold}{OT1}{cmss}{bx}{n}

  \ifAMStwofonts

    \ifCUPmtlplainloaded \else

      \DeclareSymbolFont{UPM}{U}{eur}{m}{n}

      \SetSymbolFont{UPM}{bold}{U}{eur}{b}{n}

      \DeclareSymbolFont{AMSa}{U}{msa}{m}{n}

      \DeclareMathSymbol{\upi}{0}{UPM}{"19}

      \DeclareMathSymbol{\umu}{0}{UPM}{"16}

      \DeclareMathSymbol{\upartial}{0}{UPM}{"40}

      \DeclareMathSymbol{\leqslant}{3}{AMSa}{"36}

      \DeclareMathSymbol{\geqslant}{3}{AMSa}{"3E}

      \let\leq=\leqslant \let\le=\leqslant

       \let\ge=\geqslant

    \fi

  \fi

\fi 

\ifCUPmtlplainloaded \else

  \ifAMStwofonts \else 

    \def\upi{\pi}

    \def\umu{\mu}

    \def\upartial{\partial}

  \fi

\fi

\title[A Spectropolarimetric Atlas of Seyfert 1 Galaxies]{A 
Spectropolarimetric Atlas of Seyfert 1 Galaxies}
\author[J. E. Smith et al.] {J. E. Smith,$^1$\thanks{Email: 
jsmith@star.herts.ac.uk} S. 
Young,$^1$ A. Robinson,$^1$ E. A. Corbett,$^2$ M. E. Giannuzzo,$^1$
\newauthor
D. J. Axon,$^1$ and J. H. Hough$^1$\\
$^1$Department of Physical Sciences, University of Hertfordshire, 
Hatfield, AL10 9AB, UK.\\
$^2$Anglo-Australian Observatory, PO Box 296, Epping, NSW 1710, Australia.}

\date{Accepted 2001 December 31.
      Received 2001 December 31;
      in original form 2001 December 31}

\pagerange{\pageref{firstpage}--\pageref{lastpage}}

\pubyear{2002}

\begin{document}

\maketitle

\label{firstpage}

\begin{abstract}

We present optical spectropolarimetry of the nuclei of 36 Seyfert 1 
galaxies, obtained with the William Herschel and the Anglo--Australian 
Telescopes from 1996 to 1999.  In 20 of these, the optical emission 
from the active nucleus is intrinsically polarized.  We have measured 
a significant level of polarization in a further 7 objects but these 
may be heavily contaminated by Galactic interstellar polarization.  
The intrinsically polarized Seyfert 1s exhibit a variety of 
characteristics, with the average polarization ranging from $< 0.5$ 
to 5 per cent and many showing variations in both the degree and 
position angle of polarization across the broad H$\alpha$ emission 
line.  We identify a small group of Seyfert 1s that exhibit 
polarization properties similar to those of Seyfert 2 galaxies in 
which polarized broad-lines have been discovered.  These objects 
represent direct observational evidence that a Seyfert 2-like 
far-field polar scattering region is also present in Seyfert 1s.  
Several other objects have features that can be explained in terms of 
equatorial scattering of line emission from a rotating 
disk.  We propose that much of the diversity in the polarization 
properties of Seyfert galaxies can be understood in terms of a model 
involving both equatorial and polar scattering, the relative 
importance of the two geometries as sources of polarized light being 
determined principally by the inclination of the system axis to the 
line-of-sight.

\end{abstract}


\section{Introduction}
\label{intro}

Spectropolarimetry has proved to be an important tool in the 
development of unified theories of active galactic nuclei (AGN). Its 
strength is that it provides an alternative view into the inner 
regions of the active nucleus, in addition to the direct line-of-sight. 
This allows us to probe the structure and kinematics of both 
the polarizing material and the emission source.

The discovery of polarized broad-line emission from the Seyfert 2 
galaxy NGC\,1068 (Antonucci \& Miller 1985), prompted the suggestion 
that Seyfert Types 1 and 2 nuclei are intrinsically the same class of 
object viewed at different orientations.  In the case of Seyfert 2s 
our direct view of the continuum source and broad emission line region 
(BLR) is blocked by an optically and geometrically thick torus.  
Antonucci \& Miller (1985) proposed that the optical polarized flux 
spectrum of NGC\,1068 can be explained as scattering by free electrons 
above the poles of the torus.  Subsequent modelling (Miller, Goodrich 
\& Matthews 1991; Young et al.  1995) has confirmed that scattering 
by free electrons in a conical region along the poles of the torus can 
successfully explain the observed optical and ultraviolet polarization 
characteristics of this source.

This basic polar scattering picture seems to be generally valid for 
Seyfert 2 nuclei.  Since scattered light has an electric vector (${\bf 
E}$ vector) position angle (PA) perpendicular to the scattering plane 
(the plane containing the incident and scattered ray), the polarized 
radiation has an ${\bf E}$ vector perpendicular to the scattering cone 
axis.  It seems reasonable to suppose that the principal axis of the 
system is defined by the rotation axis of the accretion disk and that 
the radio source axis, the torus and the scattering cone are 
co-aligned with this axis.  Scattered light will then be polarized 
with its ${\bf E}$ vector perpendicular to the axis of the radio 
source.  Observations support this picture: in Type 2 Seyferts the 
optical polarization PA is almost always perpendicular to the 
projected radio source axis (e.g., Antonucci 1983; Brindle et al.  
1990).

In contrast, the optical polarization properties of Type 1 Seyfert 
galaxies are not consistent with simple polar scattering.  According 
to the Seyfert unified scheme Type 1's are viewed much closer to the 
system axis than Type 2's.  Nevertheless, if only polar scatterers are 
present, Type 1's should still be polarized perpendicular to the 
projected radio source axis.  In fact, the optical polarization ${\bf 
E}$ vector is more often {\em aligned} with this axis (Antonucci 1983, 
1984, 2001; Martel 1996; M96). This implies that scattered light emerging 
from Type 1 nuclei follows a different path to that in Type 2's, 
suggesting, in turn, that the simplest unification model geometry 
including only a single, polar scattering `mirror', is incomplete.  
Evidently, we require an additional source of scattered light in 
Seyfert 1s in order to explain the alignment of the polarized light 
${\bf E}$ vectors with their radio axes.  The location, structure and 
kinematics of this source are presently unknown.  It is possible that 
in Seyfert 2s this component is hidden from view by the postulated 
torus or is entirely absent, though the latter case would conflict 
with the unified theory.  Another possibility is that both scattering 
components are viewed directly but only one dominates the observed 
polarization in any given object, since the amount of scattered 
radiation that is polarized is aspect dependent.

The observations presented here are intended to provide a data set 
with which we can investigate the origin of the optical polarization 
in Type 1 Seyfert galaxies.  There have been relatively few 
spectropolarimetric observations of Seyfert 1s in comparison to the 
extensive studies of Seyfert 2s.  The first high signal-to-noise 
spectropolarimetric observations of Type 1 nuclei were carried out by 
Goodrich \& Miller (1994; GM94).  These authors studied the 
polarization of the broad H$\alpha$ line and found a diversity of 
characteristics, including PA rotations across the line profile.  They 
also pointed out that scattering in an optically thin disk, in the 
equatorial plane of the system, can produce optical polarization PA's 
aligned with the radio axis, assuming that the latter is co-axial with 
the disk.  M96 analysed the optical polarization of 7 Seyfert 1s by 
decomposing the total flux broad H$\alpha$ line into several different 
velocity components and searching for geometrical relationships 
between the inferred polarizations of these components and 
morphological and dynamical features of the host galaxy.  More 
recently, Schmid et al.  (2000, 2001) have presented ESO VLT 
spectropolarimetry of 5 Seyfert 1 galaxies, including 3 that are also 
included in our sample.  We have previously presented optical 
spectropolarimetric observations of Mrk\,509 at 3 epochs, obtained as 
part of this programme (Young et al.  1999).  We show that the 
polarization of the broad H$\alpha$ line is variable on relatively 
short timescales, indicating that some of the scattering occurs in a 
compact region.  The variable component also has an ${\bf E}$ vector 
that is parallel to the small scale radio axis, consistent with an 
origin in a compact scattering region in the equatorial plane of the 
system.  Further modelling, Young (2000), shows that altering the flux 
incident upon the scattering regions can produce the time variable 
polarization observed.

In this paper we present spectropolarimetric observations of 36 Type 1 
Seyfert nuclei, obtained at the William Herschel telescope (WHT) and 
the Anglo--Australian telescope (AAT) between 1996 and 1999.  In 
Section~\ref{obs} the observations and data reduction procedures are 
described.  The results are presented in Section~\ref{results} along 
with brief descriptions of the properties of the individual galaxies.  
In Section~\ref{discussion} we review the results and discuss their 
implications for the scattering regions, the BLR and the Seyfert 
unification scheme.

\section[]{Observations and Data Reduction}
\label{obs}

A log of all observations is presented in Table~\ref{tab:log}.

\subsection{AAT Observations}
\label{aat-obs}

Observations were made at the 3.9-m AAT on the nights of 1996 August 
24; 1997 May 30 \& 31; 1997 August 2, 3 \& 4. The Royal Greenwich 
Observatory spectrograph was used in conjunction with the waveplate 
modulator and a 1024$\times$1024 pixel Tektronix CCD detector. The 
270R (270 lines/mm) grating was used giving a wavelength range of 
$\approx$ 3460$\mathrm{\AA}$ allowing simultaneous coverage of H$\alpha$ 
and H$\beta$. The dispersion was 3.4$\mathrm{\AA}$ pixel$^{-1}$. The 
slit width for the 1996 August observations was 1.5$\arcsec$ and for 
the remaining observations it was 2.0$\arcsec$. A two-hole dekker 
with 2.7$\times$2.7 arcsec$^{2}$ apertures separated by 27$\arcsec$ 
was used, with the object located in one aperture and the other 
providing a sky measurement.

\subsection{WHT Observations}
\label{wht-obs}

Observations were made at the 4.2-m WHT on the nights of 1997 February 
8 \& 10; 1997 June 29; 1998 September 30 \& October 1 \& 2; 1999 June 
17 \& 18.  The red arm of the ISIS dual beam spectrograph was used in 
standard polarimetry mode with a 1024$\times$1024 pixel Tektronix CCD 
detector.  For all observations, the R316R (316 lines/mm) grating was 
used, giving a wavelength range of $\approx$ 1500$\mathrm{\AA}$ which, 
when centered on the H$\alpha$ line, allows coverage of its extended 
wings and the adjacent continuum.  The R316R grating has a dispersion 
of 1.5$\mathrm{\AA}$ pixel$^{-1}$.  The slit width for 1997 February 
\& June was 1.0$\arcsec$, for the remaining observations it was 
1.2$\arcsec$.  A comb dekker was used to prevent overlapping of the 2 
sets of spectra produced by the beam-splitting calcite slab.  The 
dekker apertures were 2.7$\arcsec$ and separated by 18$\arcsec$.\\

\begin{table*}

\centering


\caption{Summary log of the observations. The exposure times are the 
sum of equal exposures at each of the 4 waveplate positions.}

\begin{tabular}{@{}lccrr@{}}

\hline

   Object     &    z    &   Telescope   &   Date  & Exposure          
\\

              &         &               &         &    (seconds)      
\\

\hline

 Akn 120      &  0.032  &     WHT       & 10/2/97 &   8$\times$1000   
\\

	      &         &               & 30/9/98 &   4000            \\

	      & 	&		& 1/10/98 &   4000            \\

              &		&		& 	  &   2400            \\

 Akn 564      &  0.025  &     WHT       & 30/9/98 &   4000            
\\

	      &         &               & 2/10/98 &   3600            \\

 ESO 012$-$G21  &  0.030  &     AAT       & 4/8/97  &   6$\times$1200   
\\

 ESO 113$-$IG45 &  0.047  &     AAT       & 3/8/97  &   2$\times$1200   
\\

 ESO 141$-$G35  &  0.016  &     AAT       & 30/5/97 &   6$\times$1200   
\\

 Fairall 51   &  0.014  &     AAT       &  4/8/97 &   4$\times$1200   
\\

 IZw1         &  0.061  &     AAT       & 2/8/97  &   6$\times$1200   
\\

	      &         &     WHT       & 30/9/98 &   2$\times$4000   \\

    	      &	  	&	        & 2/10/98 &   2$\times$2400   \\

KUV 18217+6419& 0.297   &     WHT       & 1/10/98 &   4000            
\\

              &         &               & 2/10/98 &   2$\times$4000   
\\

 Mrk 6        &  0.018  &     WHT       & 8/2/97  &   6$\times$1600   
\\

	      &         &               & 2/10/98 &   4800   \\

 Mrk 279      &  0.029  &     WHT       & 10/2/97 &   6$\times$1000   
\\

 Mrk 290      & 0.030   &     WHT       & 17/6/99 &   4000	      \\

 	      &         &               & 18/6/99 &   2800            \\

 Mrk 304      & 0.066   &     AAT       & 30/5/97 &   4$\times$1200   
\\

	      &         &               &         &   4$\times$960    \\

 Mrk 335      & 0.026   &     WHT       & 2/10/98 &   3600            
\\

 	      &         &               &         &   2400            \\

 Mrk 509      &  0.034  &     AAT       & 24/8/96 &   6$\times$1000   
\\

              &         &               & 31/5/97 &   4$\times$1200   
\\

	      &         &               & 2/8/97  &   4$\times$1200   \\ 

 Mrk 705      &  0.029  &     WHT       & 10/2/97 &   6$\times$1200   
\\ 

 Mrk 841      &	0.036 	&     AAT       & 30/5/97 &   8$\times$1200   
\\ 

 Mrk 871      & 0.032   &     AAT       & 31/5/97 &   8$\times$1200   
\\

 Mrk 876      & 0.129   &     WHT       & 29/6/97 &   4000            
\\

 Mrk 896      & 0.026   &     AAT       & 31/5/97 &   8$\times$1200   
\\

 Mrk 915      & 0.024   &     AAT       & 4/8/97  &   8$\times$1200   
\\

 Mrk 926      & 0.047   &     AAT       & 2/8/97  &   6$\times$1200   
\\	                                                                  

 Mrk 985      & 0.031   &     AAT       & 24/8/96 &   8$\times$1600   
\\

MS 1849.2$-$7832& 0.042   &     AAT       & 4/8/97  &   8$\times$1200   
\\

 NGC 3516     & 0.009   &     WHT       & 10/2/97 &   4$\times$1600   
\\

 NGC 3783     & 0.010   &     AAT       & 30/5/97 &   6$\times$1200   
\\

 NGC 4051     & 0.002   &     WHT       & 17/6/99 &   2800            
\\

 NGC 4593     & 0.009   &     WHT       & 10/2/97 &   4$\times$1600   
\\

	      &         &     AAT       & 31/5/97 &   6$\times$1200   \\

 NGC 5548     & 0.017   &     WHT       & 29/6/97 &   400	      \\

              &         & 	        &         &   3$\times$800    \\

 NGC 6104     & 0.028   &     WHT       & 17/6/99 &   2000            
\\

              &         &               & 18/6/99 &   2000            
\\

 NGC 6814     &  0.005  &     AAT       & 24/8/96 &   4$\times$1000   
\\

 	      &         &     WHT       & 29/6/97 &     1200          \\

 	      &         &               &         &     2400          \\

 NGC 7213     & 0.006   &     AAT       & 31/5/97 &   4$\times$720    
\\

 NGC 7469     & 0.016   &     WHT       & 29/6/97 &   400             
\\

	      &         &               &         &   2400            \\

              &         &     AAT       & 3/8/97  &   4$\times$1200   
\\

 	      &         &     WHT       & 30/9/98 &   2$\times$4000   \\

 NGC 7603     & 0.030   &     WHT       & 29/6/97 &   2400            
\\

 PG 1211+143  & 0.081   &     WHT       & 8/2/97  &   3$\times$1600   
\\

 UGC 3478     & 0.013   &     WHT       & 2/10/98 &   1200            
\\

 WAS 45       & 0.025   &     WHT       & 18/6/99 &   4000            
\\

\hline

\end{tabular}


\label{tab:log}
\end{table*}

\subsection{Data reduction and calibration}
\label{redcal}

The target sources and standard stars were observed using the standard 
spectropolarimetry procedure of taking equal exposures at half-wave 
plate angles of 0, 22.5, 45 and 67\degr.

Polarized standards were observed to determine the zero point of the 
polarization PA. Unpolarized standards were observed to check on the 
functioning of the system and the instrumental polarization. 
Observations of spectrophotometric standards were obtained to allow 
flux calibration of our spectra and the removal of atmospheric 
absorption features. Wavelength calibration was achieved via 
observations of CuAr or CuNe lamps.

Data reduction was performed using standard techniques within the 
Starlink packages {\sc tsp} (Bailey 1997) and {\sc figaro} 
(Shortridge et al. 1999). The relative intensities of the orthogonally 
polarized o and e rays are measured for each half-wave plate angle 
and combined using the {\sc tsp} routine {\sc ccd2pol} to obtain the 
Stokes I, Q and U parameters in each wavelength bin.

In most cases, multiple observations were made of each target.  Those 
taken during the same observing run were combined to form average I, Q 
and U spectra from which the degree and position angle of polarization 
(hereafter denoted $p$ and $\theta$, respectively) were calculated for 
each wavelength bin.  We have also obtained observations of 7 objects 
during more than one observing run.

The average values of $p$ and $\theta$ for both the continuum and 
broad H$\alpha$ line are listed in Table~\ref{tab:res}.  Both $p$ 
and $\theta$ were calculated using the {\sc ptheta} routine in the 
Starlink package {\sc polmap} (Harries 1996). This sums the Stokes 
parameters over user-defined wavelength bins and then calculates
the average $p$ and $\theta$ with their corresponding errors. For sources 
observed during different observing runs we list the measured polarizations 
at each epoch. The continuum polarization was measured from wavelength 
regions, both bluewards and redwards of H$\alpha$, considered to be 
uncontaminated by strong emission-lines. The exception is Mrk\,6, for 
which the observations obtained in 1997 February had poor 
wavelength centring, allowing us to define a continuum range only on 
the red side of H$\alpha$.  For consistency, the continuum 
polarization was also measured only on the red side of H$\alpha$ for 
the 1998 October observations.  In all cases where an object was 
observed at different epochs, the same wavelength ranges were used to 
measure the continuum polarization.

The average broad H$\alpha$ polarization was measured, over the 
wavelength range given in Table~\ref{tab:res}, after first subtracting the 
continuum.  Continuum subtraction can be carried out in a number of 
ways.  One is to subtract the average, normalized continuum Q and U 
parameters from the line complex.  This not only removes the continuum 
contribution to the measured polarization within the line profile but 
also any similar component of emission line polarization.  This method 
was used by Young et al. (1999) to identify all the polarization 
components of the broad H$\alpha$ emission line in Mrk\,509.  Here we 
adopt a different method, which retains all of the polarized 
broad-line flux, including that which is polarized like the continuum.  
The polarization of the broad H$\alpha$ line was determined by fitting 
a first-order polynomial to the Stokes spectra (I, Q, U) using {\sc polmap}. 
This fit was then subtracted from the polarization spectra to leave the intrinsic 
polarization of the H$\alpha$ emission line complex.  In either case, 
the diluting effect of starlight from the host galaxy is removed. The 
polarization spectra shown in Figs.~\ref{fig:a120}--\ref{fig:was45} have not 
been corrected in this manner and therefore retain a polarized continuum and 
starlight contribution. However, Figs.~\ref{fig:m6cs} and Figs.~\ref{fig:a120cs} 
show the continuum-subtracted plots for Mrk 6 and Akn 120 respectively. For 4 
objects, ESO\,113$-$IG45, NGC\,3516, NGC\,4593 (1997 February) and 
NGC\,6104, the data do not have sufficient signal-to-noise to 
accurately subtract the continuum and hence it was not possible to 
measure the intrinsic broad H$\alpha$ polarization.

\begin{table*}

\centering


\caption{Measured average polarizations and other properties of the 
objects as described in Section~\ref{redcal}.}

\begin{tabular}{@{}rccccllcccc@{}}

\hline

Object & Observing &Broad H$\alpha$ & \multicolumn{2}{c} {$p$}  &  
\multicolumn{2}{c} {$\theta$} & Galactic& 
\multicolumn{2}{c}{Interstellar P}\\

       &    Run    & Range ($\mathrm{\AA}$)  & Cont.      & 
H$\alpha$  &  Cont.     & H$\alpha$         & E(B -- V)  &  
typical      &        max        \\

       &           &                & (per cent)       & (per 
cent)       &  ($^\circ$)& ($^\circ$)        & (mag)   &  (per 
cent)         &    (per cent)           \\

\hline

Akn 120 	&Feb 97 &6600-7000&0.79$\pm$0.04&0.26$\pm$0.05&70.9$\pm$1.3 
& 64.9$\pm$5.3& 0.128  & 0.38 & 1.15 &\\

        	&Oct 98 &         &0.35$\pm$0.01&0.40$\pm$0.02&77.0$\pm$1.1 
& 74.9$\pm$1.2&        &      &      &\\

Akn564   	&Oct 98 &6650-6800&0.52$\pm$0.02&0.34$\pm$0.05&87.0$\pm$1.3 
& 92.5$\pm$3.9& 0.060  & 0.18 & 0.54 &\\

ESO 012$-$G21   	&Aug 97	
&6710-6860&0.41$\pm$0.05&0.57$\pm$0.14&96.5$\pm$3.6 & 89.9$\pm$7.2& 
0.080  & 0.24 & 0.72 &\\

ESO 113$-$IG45    &Aug 97 &6650-7100&0.37$\pm$0.13&             
&45.0$\pm$10.9&             & 0.027  & 0.08 & 0.24 &\\

ESO 141$-$G35     &May 97 
&6550-7050&1.30$\pm$0.03&0.97$\pm$0.06&179.1$\pm$0.6&179.9$\pm$1.8& 
0.054  & 0.16 & 0.49 &\\

Fairall 51      &Aug 97 
&6500-6800&4.12$\pm$0.03&5.19$\pm$0.07&141.2$\pm$0.2&138.0$\pm$0.4& 
0.108  & 0.32 & 0.97 &\\

IZw1    	&Aug 97 
&6875-7050&0.72$\pm$0.04&0.30$\pm$0.08&146.1$\pm$1.7&132.6$\pm$7.6& 
0.065  & 0.20 & 0.60 &\\

        	&Oct 98 &         
&0.67$\pm$0.01&0.31$\pm$0.02&151.6$\pm$0.5&133.8$\pm$1.9&        
&      &      &\\

KUV 18217+6419  &Oct 98 
&8350-8800&0.28$\pm$0.01&0.12$\pm$0.02&143.3$\pm$2.0&156.5$\pm$5.5& 
0.043  & 0.13 & 0.39 &\\

Mrk 6   	&Feb 
97&6500-6900&0.90$\pm$0.03&0.85$\pm$0.04&132.4$\pm$1.1&130.0$\pm$1.2& 
0.136  & 0.41 & 1.22 &\\

        	&Oct 98&         
&0.90$\pm$0.02&0.86$\pm$0.03&156.5$\pm$0.8&155.2$\pm$0.9&        
&      &      &\\

	      	&      &         &	       &             &		   &	         &	  
&	 &      &\\

Mrk 279       	&Feb 
97&6640-6880&0.48$\pm$0.04&0.20$\pm$0.06&58.9$\pm$2.4 &120.3$\pm$7.8& 
0.016  & 0.05 & 0.14 &\\

Mrk 290       	&Jun 
99&6600-6900&0.90$\pm$0.04&0.40$\pm$0.05&157.5$\pm$1.1&154.5$\pm$3.8& 
0.015  & 0.05 & 0.14 &\\

Mrk 304       	&May 
97&6700-7200&0.51$\pm$0.04&0.72$\pm$0.07&135.5$\pm$2.5&123.3$\pm$2.9& 
0.073  & 0.22 & 0.66 &\\

Mrk 335       	&Oct 
98&6600-6850&0.28$\pm$0.01&0.52$\pm$0.02&113.6$\pm$1.7&104.5$\pm$1.2& 
0.035  & 0.11 & 0.32 &\\

Mrk 509 	&Aug 
96&6600-7000&0.85$\pm$0.03&0.46$\pm$0.03&151.8$\pm$0.9&145.3$\pm$1.6& 
0.057  & 0.17 & 0.51 &\\

        	&May 97&         
&0.70$\pm$0.04&0.41$\pm$0.03&141.4$\pm$1.5&148.2$\pm$2.4&        
&      &      &\\

        	&Aug 97&         
&0.55$\pm$0.03&0.46$\pm$0.04&139.0$\pm$2.0&145.0$\pm$2.5&        
&      &      &\\

Mrk 705      	&Feb 97 
&6685-6850&0.46$\pm$0.07&0.41$\pm$0.11&49.3$\pm$6.5 & 84.0$\pm$7.5& 
0.041  & 0.12 & 0.37 &\\

Mrk 841       	&May 
97&6640-7000&1.00$\pm$0.03&0.36$\pm$0.06&103.4$\pm$1.0& 98.0$\pm$4.6& 
0.030  & 0.09 & 0.27 &\\

Mrk 871       	&May 
97&6600-7000&0.30$\pm$0.04&0.65$\pm$0.13&84.2$\pm$3.7 & 90.3$\pm$5.8& 
0.055  & 0.17 & 0.50 &\\

Mrk 876         &Jun 
97&7050-7740&0.81$\pm$0.04&0.36$\pm$0.06&110.5$\pm$1.4&112.4$\pm$4.6& 
0.027  & 0.08 & 0.24 &\\

Mrk 896       	&May 
97&6600-6800&0.27$\pm$0.03&0.23$\pm$0.11&62.0$\pm$3.0 &66.9$\pm$13.0& 
0.045  & 0.14 & 0.40 &\\

	      	&      &	&	      &		    &		  &	        &	 &	&      &\\

Mrk 915         &Aug 
97&6580-6860&0.25$\pm$0.03&0.47$\pm$0.07&105.4$\pm$3.7& 94.7$\pm$3.9& 
0.063  & 0.19 & 0.57 &\\

Mrk 926 	&Aug 97&6550-7070&0.18$\pm$0.04&0.20$\pm$0.06&44.8$\pm$6.4 & 
31.8$\pm$8.8& 0.042  & 0.13 & 0.38 &\\

Mrk 985       	&Aug 
96&6500-7100&1.12$\pm$0.02&0.61$\pm$0.03&114.3$\pm$0.4& 98.8$\pm$1.5& 
0.048  & 0.14 & 0.43 &\\

MS 1849.2$-$7832 	&Aug 
97&6680-6980&1.64$\pm$0.06&1.45$\pm$0.16&7.0$\pm$1.1  & 19.4$\pm$3.1& 
0.152  & 0.46 & 1.37 &\\

NGC 3516      	&Feb 97&6500-6740&0.15$\pm$0.04&             
&30.1$\pm$8.0  &             & 0.042  & 0.13 & 0.39 &\\

NGC 3783      	&May 
97&6500-6760&0.52$\pm$0.02&0.22$\pm$0.03&135.5$\pm$1.0 
&115.8$\pm$3.7& 0.119  & 0.36 & 1.07 &\\

NGC 4051      	&Jun 
99&6520-6640&0.55$\pm$0.04&0.47$\pm$0.09&82.8$\pm$1.8  & 
79.9$\pm$4.7& 0.013  & 0.04 & 0.12 &\\

NGC 4593 	&Feb 97&6480-6800&0.14$\pm$0.05&             
&109.5$\pm$10.8&             & 0.025  & 0.08 & 0.23 &\\

	 	&May 97&         
&0.14$\pm$0.03&0.57$\pm$0.05&149.9$\pm$5.1&143.2$\pm$2.3&        
&      &      &\\

NGC 5548      	&Jun 
97&6520-6860&0.69$\pm$0.01&0.20$\pm$0.02&33.2$\pm$0.5 & 67.3$\pm$2.4& 
0.020  & 0.06 & 0.18 &\\

NGC 6104      	&Jun 99&6560-6930&0.44$\pm$0.12&             
&119.5$\pm$8.1&             & 0.019  & 0.06 & 0.17 &\\

	      	&     &		&	      &		    &		  &	        &	 &	&      &\\

NGC 6814 	&Aug 
96&6460-6740&1.88$\pm$0.04&1.63$\pm$0.12&162.5$\pm$0.6&165.9$\pm$2.1& 
0.183  & 0.55 & 1.65 &\\

         	&Jun 97&         &1.86$\pm$0.03&1.71$\pm$0.07&2.2$\pm$0.4  
&  1.8$\pm$1.2&        &      &      &\\

NGC 7213      	&May 
97&6530-6740&0.09$\pm$0.02&0.20$\pm$0.17&146.0$\pm$7.6&79.9$\pm$23.9& 
0.015  & 0.05 & 0.14 &\\

NGC 7469 	&Jun 97&6560-6800&0.18$\pm$0.01&0.06$\pm$0.02&76.8$\pm$1.7 
&102.1$\pm$10.6& 0.069  & 0.21 & 0.62 &\\

	 	&Aug 97&         &0.22$\pm$0.03&0.04$\pm$0.05&76.2$\pm$3.0 
&101.8$\pm$34.9&        &      &      &\\

	 	&Oct 98&         &0.09$\pm$0.01&0.11$\pm$0.02&87.3$\pm$4.5 & 
95.2$\pm$6.0 &        &      &      &\\

NGC 7603      	&Jun 
97&6500-6980&0.25$\pm$0.04&0.25$\pm$0.04&132.7$\pm$1.8&136.4$\pm$4.2 
& 0.046  & 0.14 & 0.41 &\\

PG 1211+143   	&Feb 
97&6960-7220&0.27$\pm$0.04&0.11$\pm$0.06&137.7$\pm$4.5&112.1$\pm$15.8& 
0.035  & 0.11 & 0.32 &\\

UGC 3478      	&Oct 
98&6580-6710&0.84$\pm$0.06&1.10$\pm$0.10&22.1$\pm$2.2 & 19.7$\pm$2.6 
& 0.092  & 0.28 & 0.83 &\\

WAS 45        	&Jun 
99&6660-6830&0.77$\pm$0.06&2.46$\pm$0.18&165.1$\pm$2.0&144.0$\pm$2.2 
& 0.018  & 0.05 & 0.16 &\\

\hline

\end{tabular}

\label{tab:res}
\end{table*}

\section[]{Results}
\label{results}

\begin{figure}
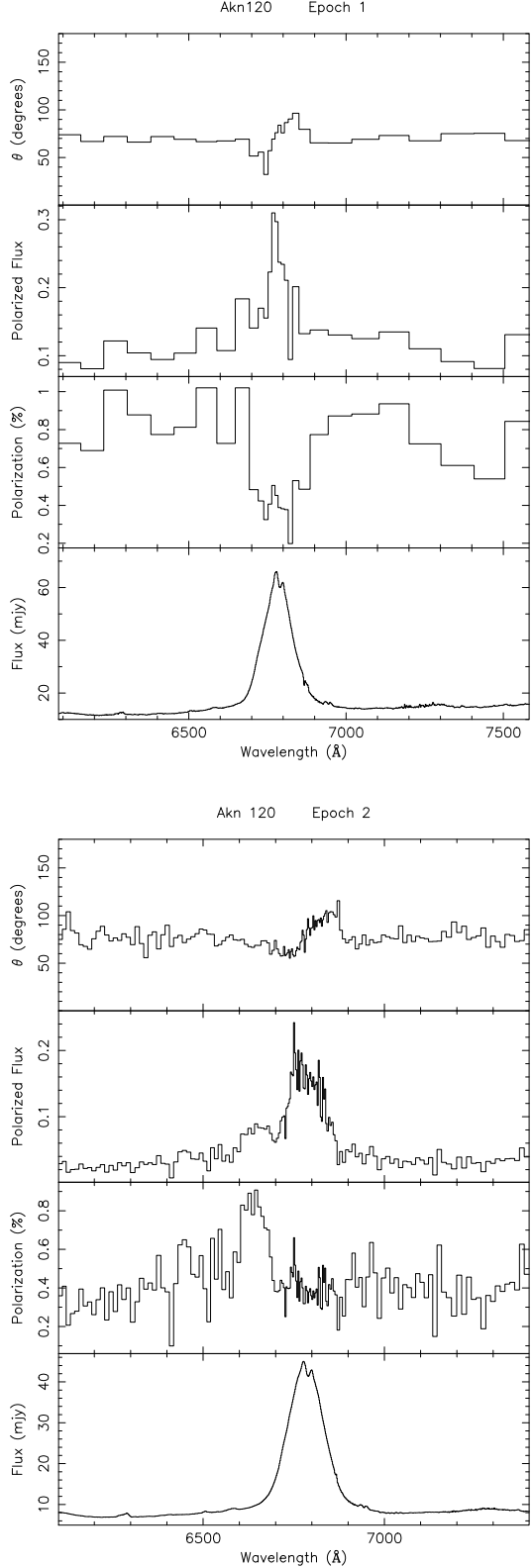


\psfig{figure=./a120_e1.eps,height=4.25in}

\psfig{figure=./a120_e2.eps,height=4.25in}

\centering{\caption{Spectropolarimetric data for Akn\,120.  In each 
frame, the panels show, from the bottom, the total flux density, the 
percentage polarization, the polarized flux density and the position 
angle of polarization ($\theta$).  The two frames show the data 
obtained in 1997 February (Epoch 1) and 1998 October (Epoch 2).  The 
polarization data are binned at 0.1 per cent.\label{fig:a120}}}

\end{figure}


\begin{figure}

\psfig{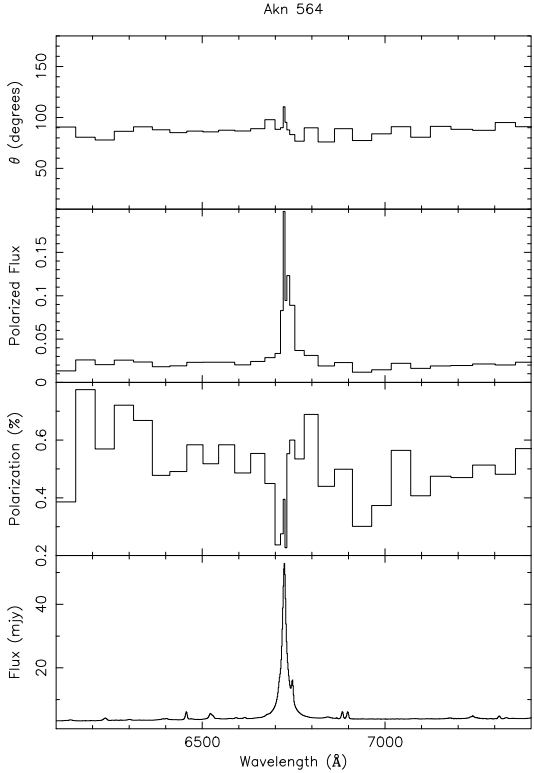}

\centering{\caption{As Fig. 1 for Akn\,564. Polarization data binned 
at 0.1 per cent.\label{fig:a564}}}

\end{figure}


\begin{figure}

\psfig{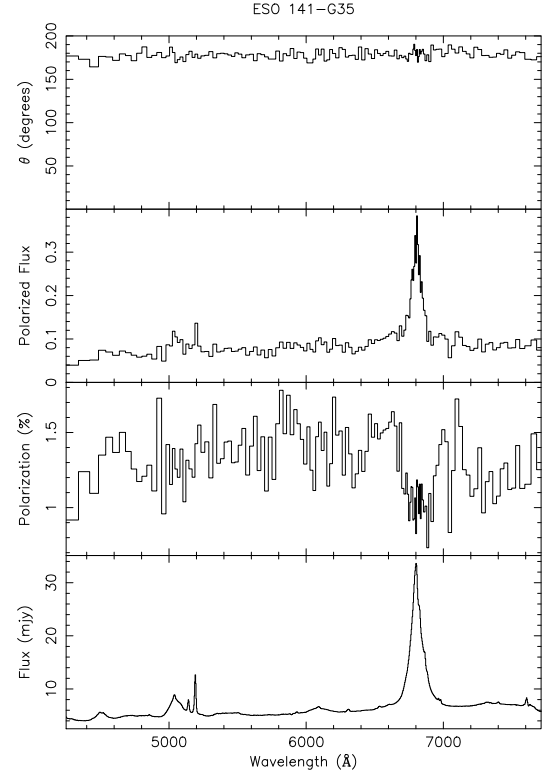}

\centering{\caption{As Fig. 1 for ESO\,141$-$G35. Polarization data 
binned at 0.2 per cent.\label{fig:e141}}}

\end{figure}


\begin{figure}

\psfig{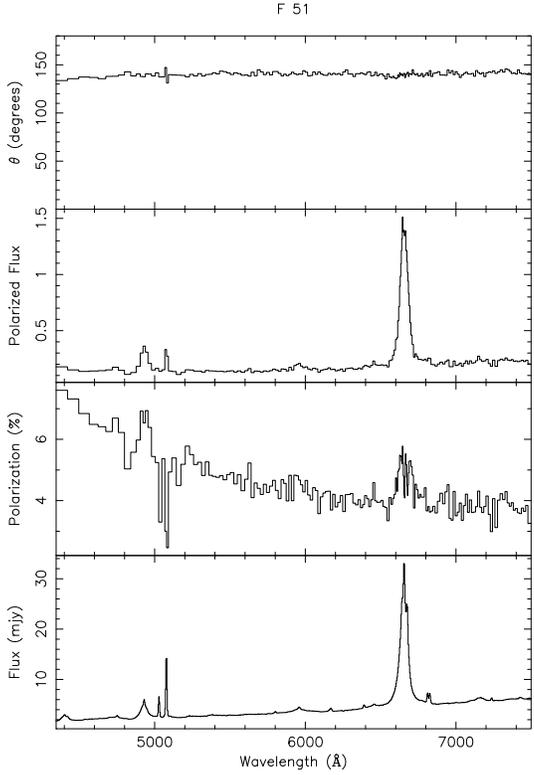}

\centering{\caption{As Fig. 1 for Fairall\,51. Polarization data 
binned at 0.3 per cent.\label{fig:f51}}}

\end{figure}


\begin{figure}

\psfig{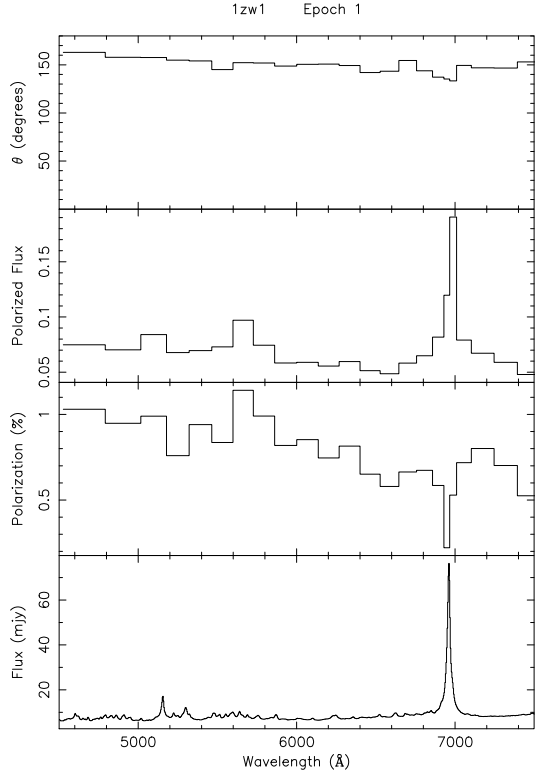}

\psfig{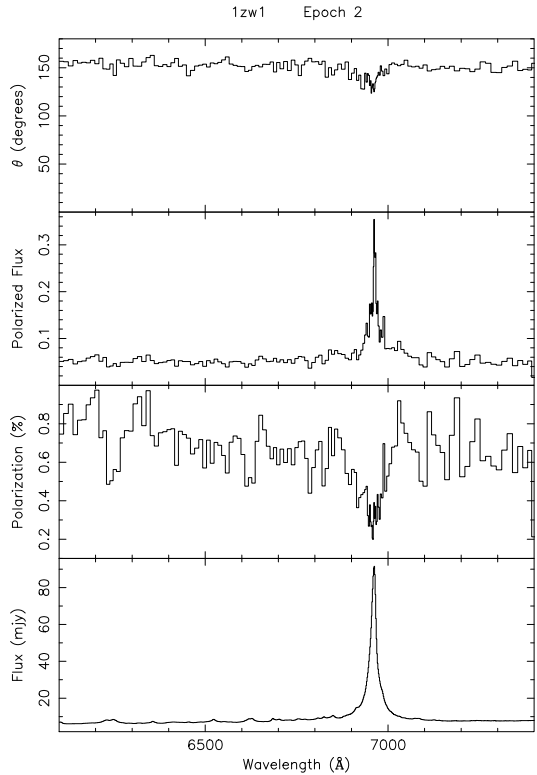}

\centering{\caption{As Fig. 1 for 1\,Zw1 -- 1997 August (Epoch 1) and  
1998 September (Epoch 2). Polarization data binned at 0.1 per cent.\label{fig:1zw1}}}

\end{figure}


\begin{figure}

\psfig{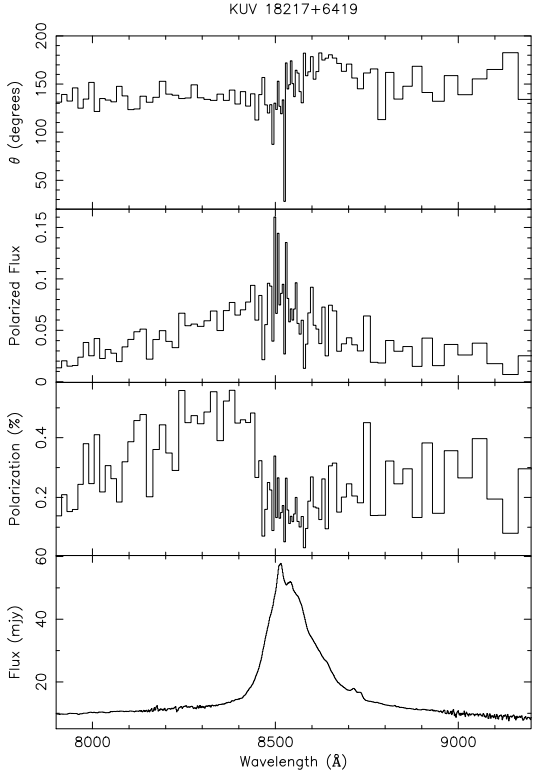}

\centering{\caption{As Fig. 1 for KUV\,18217+6419. Polarization data 
binned at 0.1 per cent.\label{fig:kuv}}}

\end{figure}


Before discussing the results, we must consider the possibility of 
contamination by foreground interstellar polarization produced in the 
Galaxy by aligned dust grains along the lines-of-sight to the target 
source.  Estimates of the line-of-sight polarization induced by the 
interstellar medium can be obtained from the empirical relationship 
between extinction, E(B$-$V), and percentage polarization established 
by Serkowski, Mathewson \& Ford (1975).  For typical conditions the 
percentage polarization induced is $\sim$ 3E(B$-$V), whilst the maximum 
observed is $\sim$ 9E(B$-$V), the latter presumably being produced by 
highly aligned dust grains.  In Table~\ref{tab:res} the E(B$-$V) from the 
reddening estimate of Schlegel, Finkbeiner \& Davis (1998) as listed 
in NED, is given for each object, along with the corresponding typical 
and maximum induced polarizations.

Interstellar polarization may also 
occur in the galaxy hosting the AGN and the effects of this are more 
difficult to assess.  However, since interstellar polarization 
varies weakly with wavelength we can conclude that the detection of 
significant PA rotations or changes in the degree of polarization 
across a relatively short wavelength range (e.g.,  across an emission 
line), indicates that the polarization is at least partly intrinsic 
to the AGN.

The unpolarized starlight from the host galaxy affects our 
measurements by diluting the intrinsic polarization of the AGN. 
Additionally, the level of starlight contribution may not be constant 
across the wavelength range of our data and can lead to an apparent 
increase in  polarization to shorter wavelengths, 
as noted in Section~\ref{polstr}. As discussed in Section~\ref{redcal}, removal of 
the continuum also removes the diluting effect of the starlight 
across the H$\alpha$ complex.

When available, published radio maps were used to determine the radio 
source PA in those objects whose radio source is extended along a 
preferred axis.  In the absence of a published measurement, the PA was 
determined by eye to a precision of $\pm10\degr$.  In several 
cases, we use PA's from unpublished maps which are quoted in the 
literature.  In Section~\ref{intpol} we give an indication, for each source, of 
the spatial scale to which the radio axis PA applies.  Angular 
distances were converted to parsecs using a Hubble constant of 75 km 
s$^{-1}$Mpc$^{-1}$.

We have measured significant intrinsic polarization in 20 objects.  
Our observations of these objects are presented in 
Section~\ref{intpol}.  The polarization spectra are shown in 
Figs.~\ref{fig:a120}--\ref{fig:was45}.  The polarized flux, percentage 
polarization and PA are binned in wavelength such that the error per 
bin in percentage polarization is constant at the values given in the 
Figure caption.  In Section~\ref{nopol} we discuss the observations of objects 
which may be significantly contaminated by interstellar polarization, 
or which are of poor signal-to-noise.

\subsection{Objects exhibiting intrinsic polarization}
\label{intpol}

\subsubsection{Akn\,120}
\label{a120}

We observed Akn\,120 at two epochs separated by $\sim$ 20 months 
(Fig.~\ref{fig:a120}), during which time a change in the polarization 
properties occured. The 1998 October observations are of better quality and show that, 
while the core and red wing of the broad H$\alpha$ profile are 
polarized at a similar level to the continuum, the blue wing exhibits 
a distinct peak in $p$.  This also appears as a blue-shifted feature 
in the polarized flux spectrum.  In our 1997 February observations 
both line wings are polarized at a similar level to the continuum but 
there is a decrease in $p$ over the line core.  During the period 
between the two observations the continuum polarization decreased by a 
factor of 2 whilst the average polarization of the broad H$\alpha$ 
line increased from 0.26$\pm$0.05 per cent to 0.40$\pm$0.02 per cent.

A PA rotation across the broad H$\alpha$ line is seen at both epochs. 
Bluewards of the line peak, $\theta$ is slightly lower 
than in the continuum but a $\sim$ 50$\degr$ rotation to 
a higher value occurs over the red side of the profile. In the 
continuum-subtracted spectra $\theta$ exhibits a swing of 
$\sim$ 80$\degr$ across the line profile (Fig.~\ref{fig:a120cs}). 

The optical polarization of Akn\,120 has also been studied by M96 and 
more recently, by Schmid et al.  (2000).  M96 observed Akn\,120 
several times between October 1993 and February 1995 and reports 
considerable variation between epochs.  He found that for the 
continuum around H$\alpha$, $p\approx 0.53$ per cent and 
$\theta\approx 68\degr$, whilst for the broad H$\alpha$ line 
itself, $p\approx 0.18$ per cent and $\theta\approx 61\degr$.  The 
polarization structure across the broad H$\alpha$ line present in our 
1997 February data is consistent with similar structure seen by M96.  
Schmid et al.  observed Akn\,120 in 1999 August, about 10 months after 
our second run.  Their results, however, are more consistent with our 
earlier observations: they measure $p \approx 0.7$ and $\theta \approx 
68\degr$ for the continuum and $p\approx 0.3$ per cent with $\theta 
\approx 56\degr$ for the broad H$\alpha$ line, respectively.  The 
$p$ and $\theta$ spectra are also broadly comparable to our 1997 
February data (although the fine structure present in Schmid et al.'s 
VLT data is not seen in our relatively noisy spectra).  The large 
change in the percentage polarization spectrum that occured between 
our 1997 February and 1998 October observations had evidently been 
reversed by the time of Schmid et al.'s observation, the spectrum 
having reverted to something close to its 1997 February form.

Condon et al. (1998) present a 1.4 GHz VLA map showing a slight 
elongation in PA $\sim 50\degr$, extending $\sim 10\arcsec$ (6.1 
kpc) from the radio core. Most of the blue side of H$\alpha$ is 
polarized at $\sim 50\degr$, approximately parallel to the radio 
axis, but most of the red side has a PA $\sim 100\degr$.

\subsubsection{Akn\,564}
\label{a564}

This source is a well-known narrow-line Seyfert 1 galaxy (NLS1).

Our spectrum (Fig.~\ref{fig:a564}) shows a decrease in $p$ in the blue 
wing of the broad H$\alpha$ line whilst the red side of the profile is 
polarized at approximately the continuum level.  There is also a 
decrease in $p$ near the line centre, which is probably due to weakly 
polarized narrow H$\alpha$.

The optical polarization of Akn\,564 has been studied by Goodrich 
(1989b), who measured a continuum polarization of $0.40\pm0.02$ per 
cent at a PA of $90.3\pm 1.5\degr$ and concluded that the 
polarization was probably interstellar in nature.  The estimated 
E(B$-$V) of 0.060 suggests that interstellar polarization may be 
significant, but the decrease in $p$ over H$\alpha$ 
indicates that this source is in fact intrinsically polarized. 

Moran (2000) presents a 3.6\,cm VLA map of Akn\,564, which shows a 
small scale `jet-like' feature extending $\sim 1\arcsec$ (480 pc) from 
the radio core in PA $\sim350\degr$.  The average values of 
$\theta$ for both the continuum and broad H$\alpha$ line differ by 
$\sim 80\degr$ from the radio axis.

\begin{figure}

\psfig{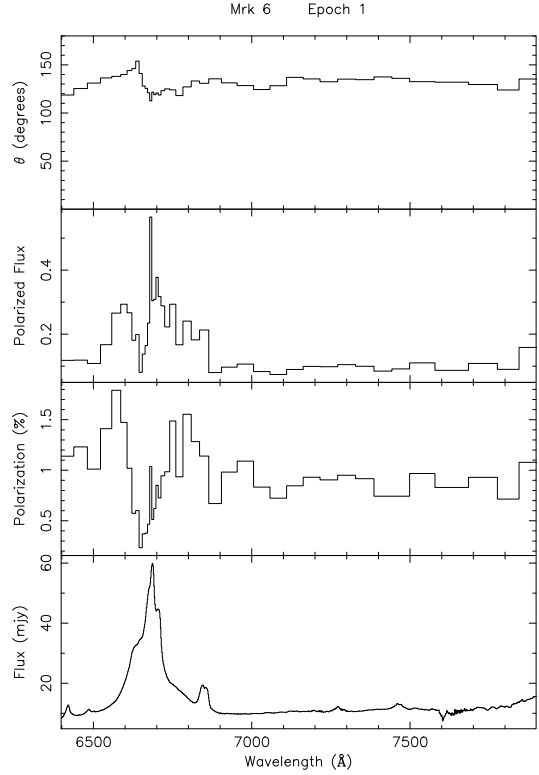}

\psfig{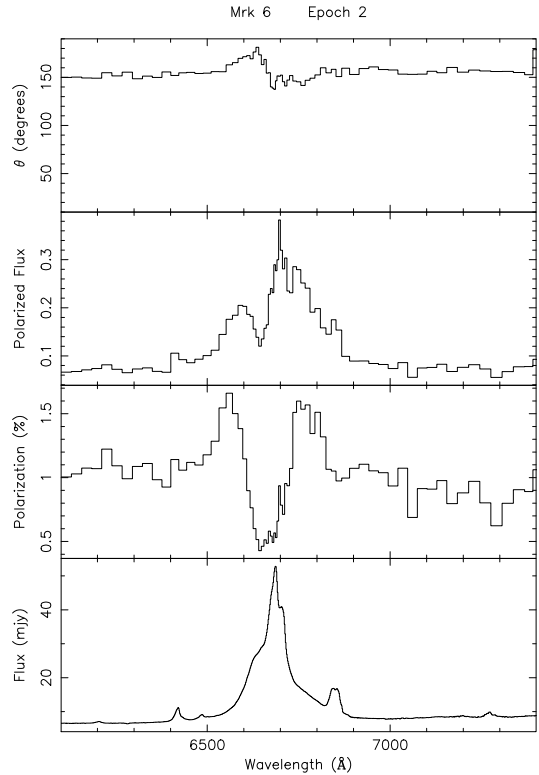}

\centering{\caption{As Fig. 1 for Mrk\,6 -- 1997 February (Epoch 1),  
1998 October (Epoch 2). Polarization data binned at 0.1 per cent.\label{fig:m6} }}

\end{figure}


\begin{figure}

\psfig{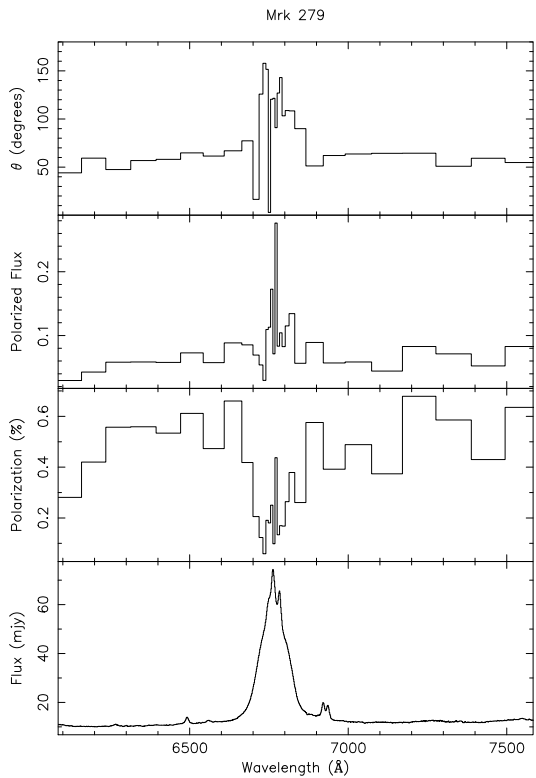}

\centering{\caption{As Fig. 1 for Mrk\,279. Polarization binned at 
0.15 per cent.\label{fig:m279}}}

\end{figure}


\begin{figure}

\psfig{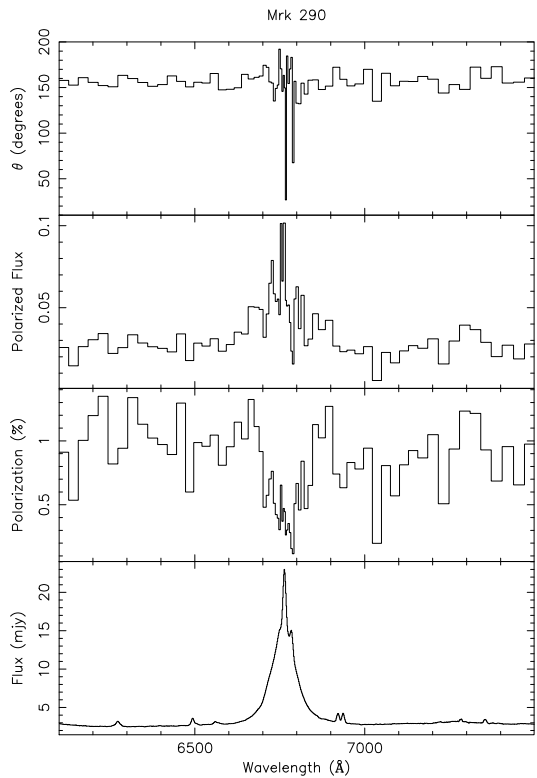}

\centering{\caption{As Fig. 1 for Mrk\,290. Polarization data binned 
at 0.2 per cent.\label{fig:m290}}}

\end{figure}


\begin{figure}

\psfig{figure=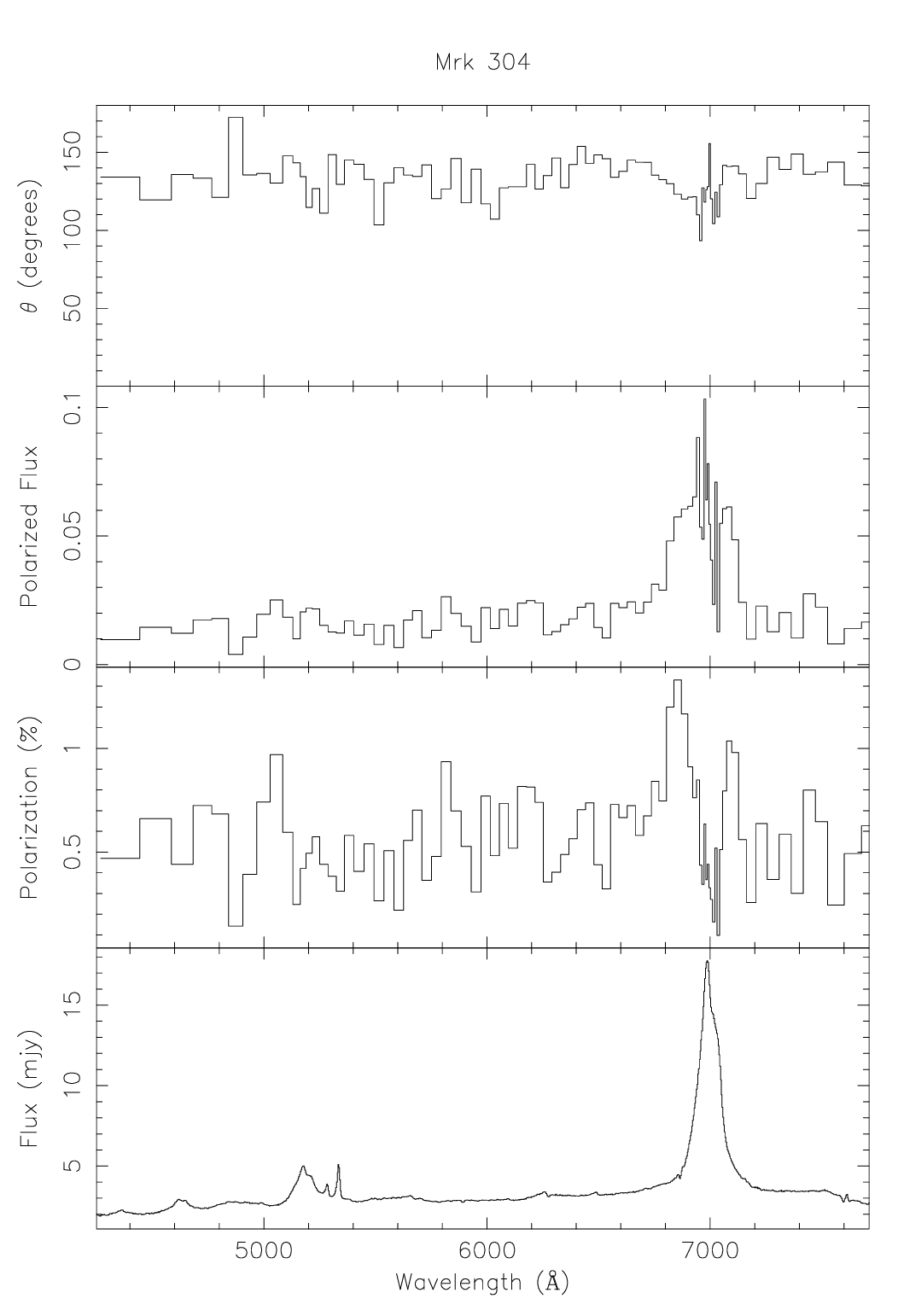,height=4.25in}

\centering{\caption{As Fig. 1 for Mrk\,304. Polarization data binned 
at 0.2 per cent.\label{fig:m304}}}

\end{figure}


\begin{figure}

\psfig{figure=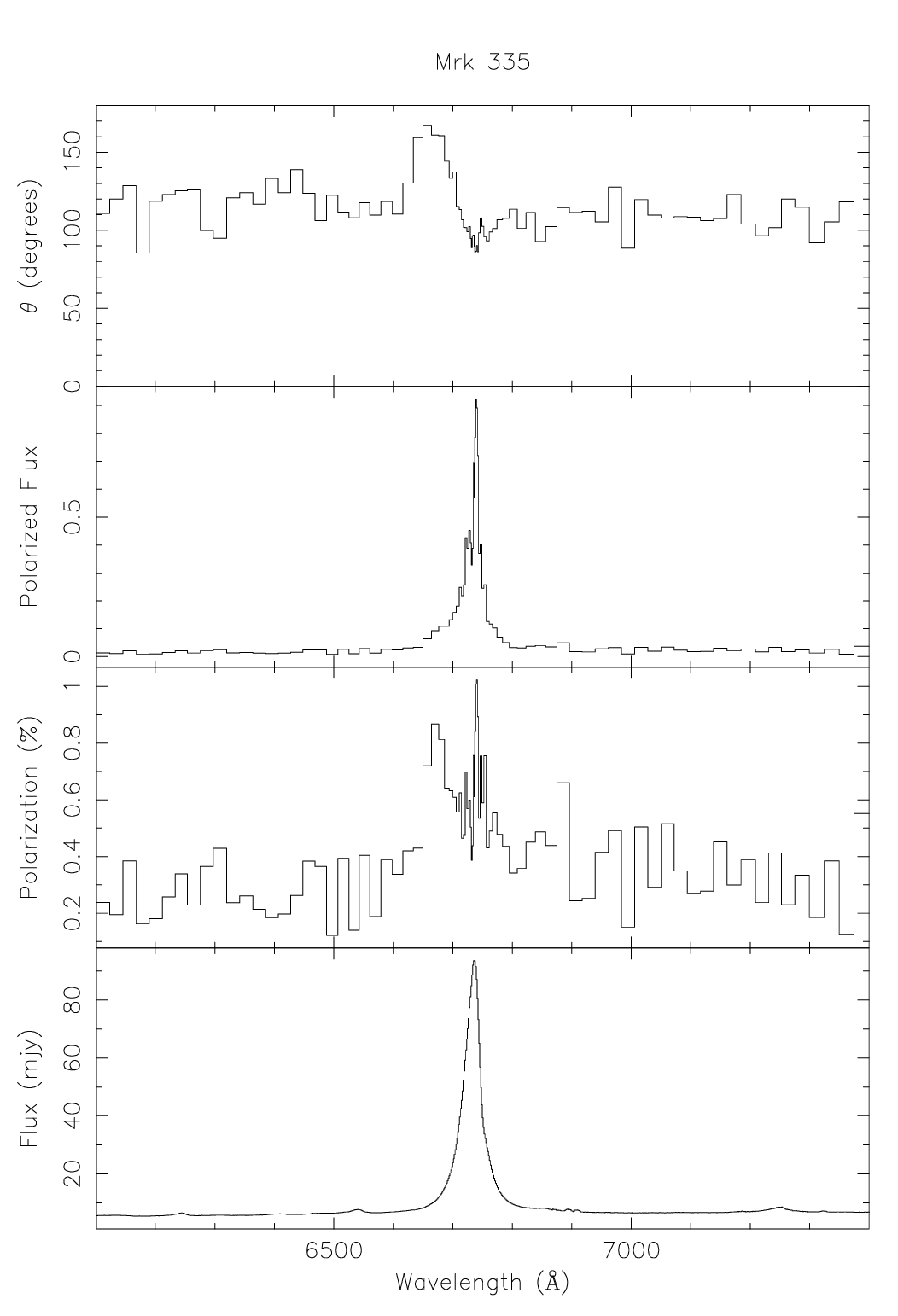,height=4.25in}

\centering{\caption{As Fig. 1 for Mrk\,335. Polarization data binned 
at 0.1 per cent.\label{fig:m335}}}

\end{figure}


\subsubsection{ESO\,141$-$G35}
\label{e141}

ESO\,141$-$G35 exhibits a constant value of $\theta$ over the spectral 
range covered by our observations (Fig.~\ref{fig:e141}).  The continuum 
is polarized at a level of $\approx 1.5$ per cent but $p$ decreases to 
a broad minimum in the core of the broad H$\alpha$ line.  There is a 
suggestion that the blue wing of broad H$\alpha$ is extended in 
polarized flux.

\subsubsection{Fairall\,51}
\label{f51}

Fairall\,51 is by far the most highly polarized source in our sample, 
exhibiting average continuum and H$\alpha$ polarizations of 4 and 5 
per cent, respectively.  Moreover, $p$ rises gradually to the blue, 
increasing from $\approx 4$ per cent near H$\alpha$, to over 7 per cent 
at 4400$\mathrm{\AA}$ (Fig.~\ref{fig:f51}).  There are also local 
increases in $p$ over both the broad H$\alpha$ and H$\beta$ lines.  
Sharp dips in $p$ are associated with the narrow [O\,{\sc iii}] lines 
and with the narrow-lines in the H$\alpha$ complex.

Schmid et al. (2001) observed this source with the VLT in 1999 August 
(2 years after our observation) and obtained very similar results.

\subsubsection{I\,Zw1}
\label{1zw1}

I\,Zw1, the archetypal NLS1, was observed at 2 epochs and 
displayed similar polarization characteristics on both occasions 
(Fig.~\ref{fig:1zw1}). 

The percentage polarization spectrum shows a decrease in $p$ over the 
broad H$\alpha$ line and in the higher signal-to-noise 1998 September 
data there is also evidence for similar decreases associated with 
emission features at observed wavelengths of $\approx$ 6229$\mathrm{\AA}$, 
6248$\mathrm{\AA}$ and 6623$\mathrm{\AA}$.  These features can be 
identified as emission-lines of He\,{\sc i} $\lambda$5876, Na\,{\sc i} 
$\lambda$$\lambda$5889.9, 5895.9 and Fe\,{\sc ii} respectively (e.g., 
Phillips 1976).

A small change in $\theta$ is evident over the broad H$\alpha$ line, 
giving it a lower average value than the continuum.  In the 
continuum-subtracted spectra there is a $\sim$ 40$\degr$ PA rotation from 
the blue to the red side of the line.

Kellerman et al. (1994) present a 6\,cm VLA radio image of I\,Zw1 which 
shows an asymmetric structure with several contours elongated at a PA 
of $\sim$ 320$\degr$ on a scale of $\sim$ 30$\arcsec$ (34 kpc).  Both 
the continuum and broad H$\alpha$ line have average polarization PA's 
approximately parallel to this elongation.

\subsubsection{KUV\,18217+6419}
\label{kuv}

This quasar is weakly polarized, but exhibits significant changes in 
percentage polarization and PA over the spectral range covered by our 
data (Fig.~\ref{fig:kuv}).

The most notable feature is a $\sim$ 70$\degr$ PA rotation on the 
red side of the broad H$\alpha$ profile.  The bulk of the line is 
polarized at a level similar to that of the continuum, but there is 
also a significant increase in $p$ associated with the far blue wing.  
As a result, the polarized flux profile appears to have an extended 
blue wing, whilst the total flux profile is asymmetric in the opposite 
sense.

Blundell \& Lacy (1995) present an 8\,GHz VLA map showing 2 
components, possibly associated with jets, straddling the radio core 
in PA $\sim 20\degr$, and further extended emission in PA 
$\sim 285\degr$.  The `jets' are separated from the core by 
$\sim 1\arcsec$ (4.95 kpc).  On larger scales, there is an additional 
feature extending $\sim 2\arcsec$ (9.9 kpc) from the core in PA 
$\sim 170\degr$.  This latter feature may be associated with the 
southern `jet' component.  The average values of $\theta$ for the 
broad H$\alpha$ line and adjacent continuum differ from the PA of the 
radio `jet' component by $\sim 40\degr$ and $\sim 57\degr$, 
respectively.  However, the average line and continuum $\theta$ values 
are similar to the PA of the larger scale radio axis.

\subsubsection{Mrk\,6}
\label{m6}

Mrk\,6 displays a highly asymmetrical broad H$\alpha$ profile and 
unusually strong and broad `narrow' lines.

We have obtained optical spectropolarimetry of Mrk 6 at two epochs 
separated by $\sim$ 20 months (Fig.~\ref{fig:m6}).  At both epochs, 
striking variations across the broad H$\alpha$ line in both degree and 
position angle of polarization are present.  In the continuum 
$p\approx 1$ per cent but rises steeply to separate peaks at about 1.5 
per cent in both the red and blue wings of the line profile.  
Conversely, $p$ drops sharply through the line core to a minimum of 
$\approx 0.5$ per cent.  In the polarized flux spectrum the H$\alpha$ 
profile has an asymmetric double-peaked structure, the red peak being 
the more prominent.  The structure in the percentage polarization 
spectra is very similar to that seen in Mrk\,509 (Young et al.  1999; 
this paper, Fig.~\ref{fig:m509}).

The polarization PA also varies significantly across the H$\alpha$ 
profile.  In the line core and red wing, $\theta$ is slightly smaller 
than in the continuum but there is a $\sim 40\degr$ rotation over 
the blue side of the profile.  In the continuum-subtracted spectra the 
amplitude of the PA swing is $\sim 70^\circ$ (Fig.~\ref{fig:m6cs}).

The average levels of polarization are similar at both observation 
epochs, as are the general forms of the structure in both the $p$ and 
$\theta$ spectra.  However, the average values of $\theta$ for the two 
epochs differ by $\sim 25^\circ$.  We believe that this shift is due 
to instrumentation problems affecting the 1997 February (first epoch) 
observations.  Therefore, in subsequent discussion we will use the 
average polarization PA's measured from the 1998 October data.

A 6\,cm MERLIN radio map presented by Capetti et al.  (1995) shows 
that the inner $\sim 0.4\arcsec$ (140\,pc) of the radio structure has 
a linear, `jet-like', morphology consisting of 3 components aligned 
along PA $\sim 170^\circ$.  Beyond $\sim 0.4\arcsec$ from the central 
radio component, the jet widens and bends to the west.  The average 
values of $\theta$ for the continuum and broad H$\alpha$ line differ 
from jet PA by $\sim 15^\circ$.  In the continuum-subtracted 
polarization spectrum the H$\alpha$ {\bf E} vector rotates from 
$\sim 130^\circ$ to $\sim 200^\circ$.  This swing in polarization PA is 
largely confined to the blue side of the line profile and is 
roughly centred on the PA of the radio axis.

\begin{figure}
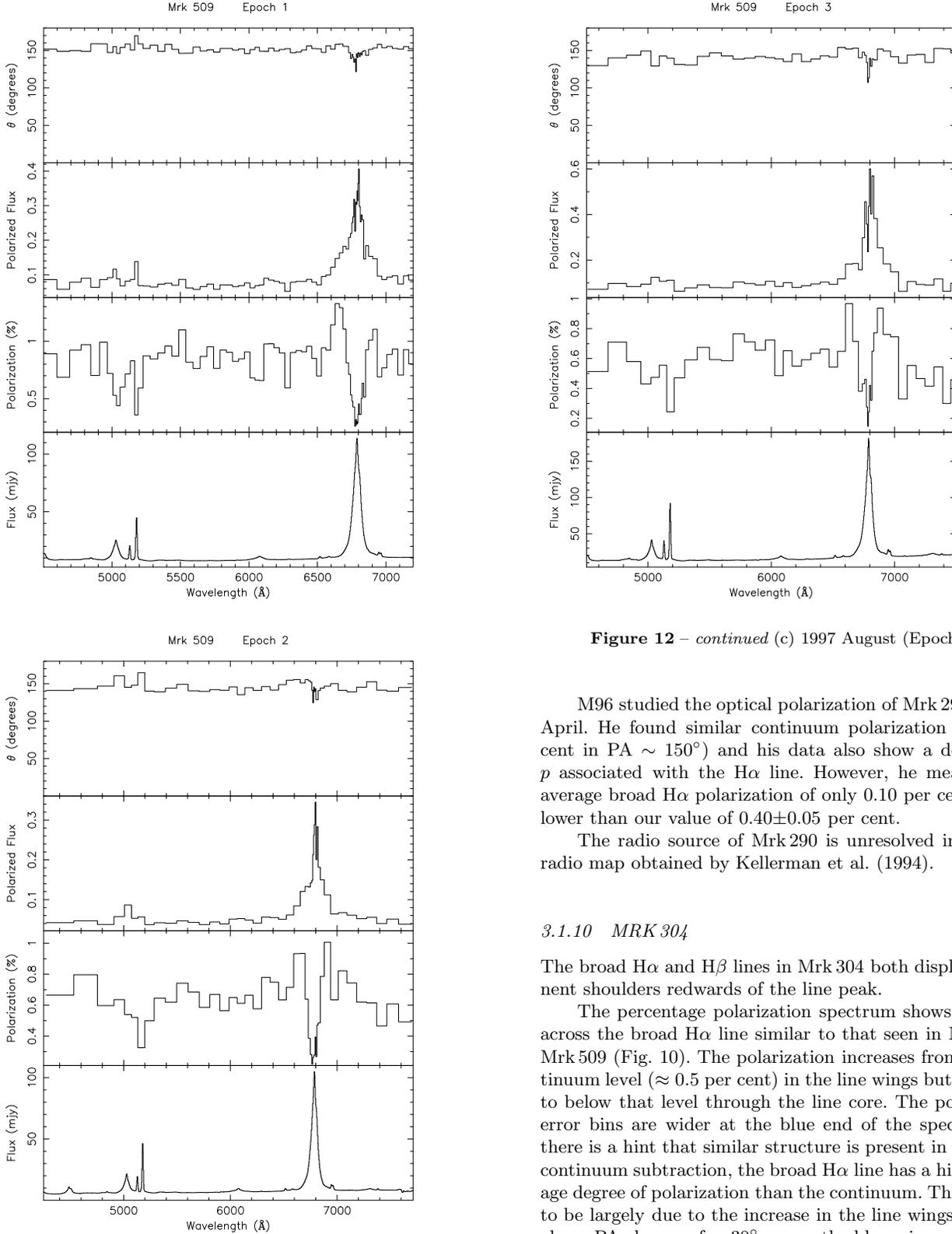


\psfig{figure=./m509_e1.eps,height=4.25in}

\psfig{figure=./m509_e2.eps,height=4.25in}

\centering{\caption{As Fig. 1 for Mrk\,509 -- 1996 August (Epoch 1),  
1997 May (Epoch 2). Polarization data binned at 0.1 per cent.\label{fig:m509}}}

\end{figure}


\begin{figure}

\psfig{figure=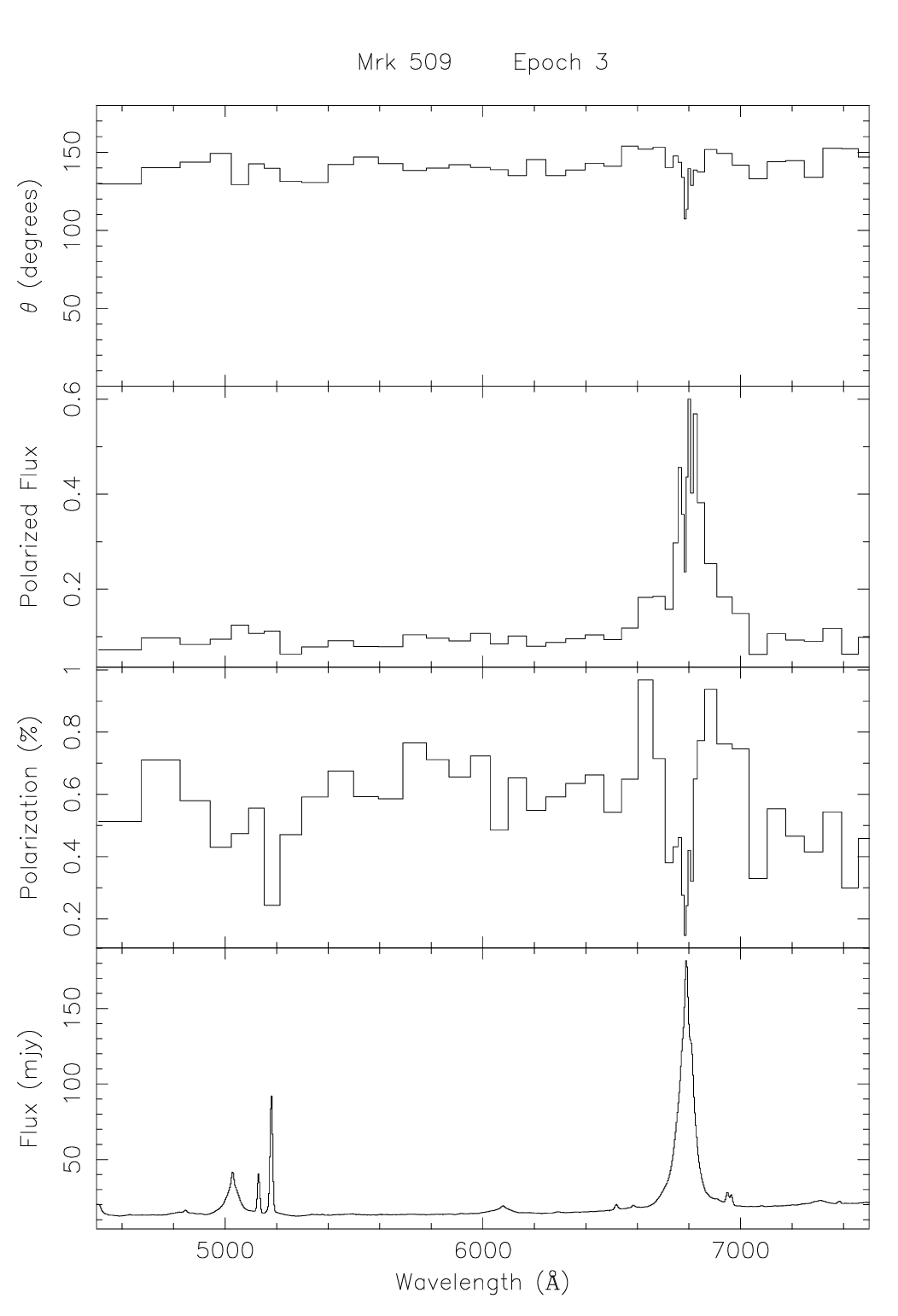,height=4.25in}

\centering{\contcaption{(c) 1997 August (Epoch 3).}}

\end{figure}


\begin{figure}

\psfig{figure=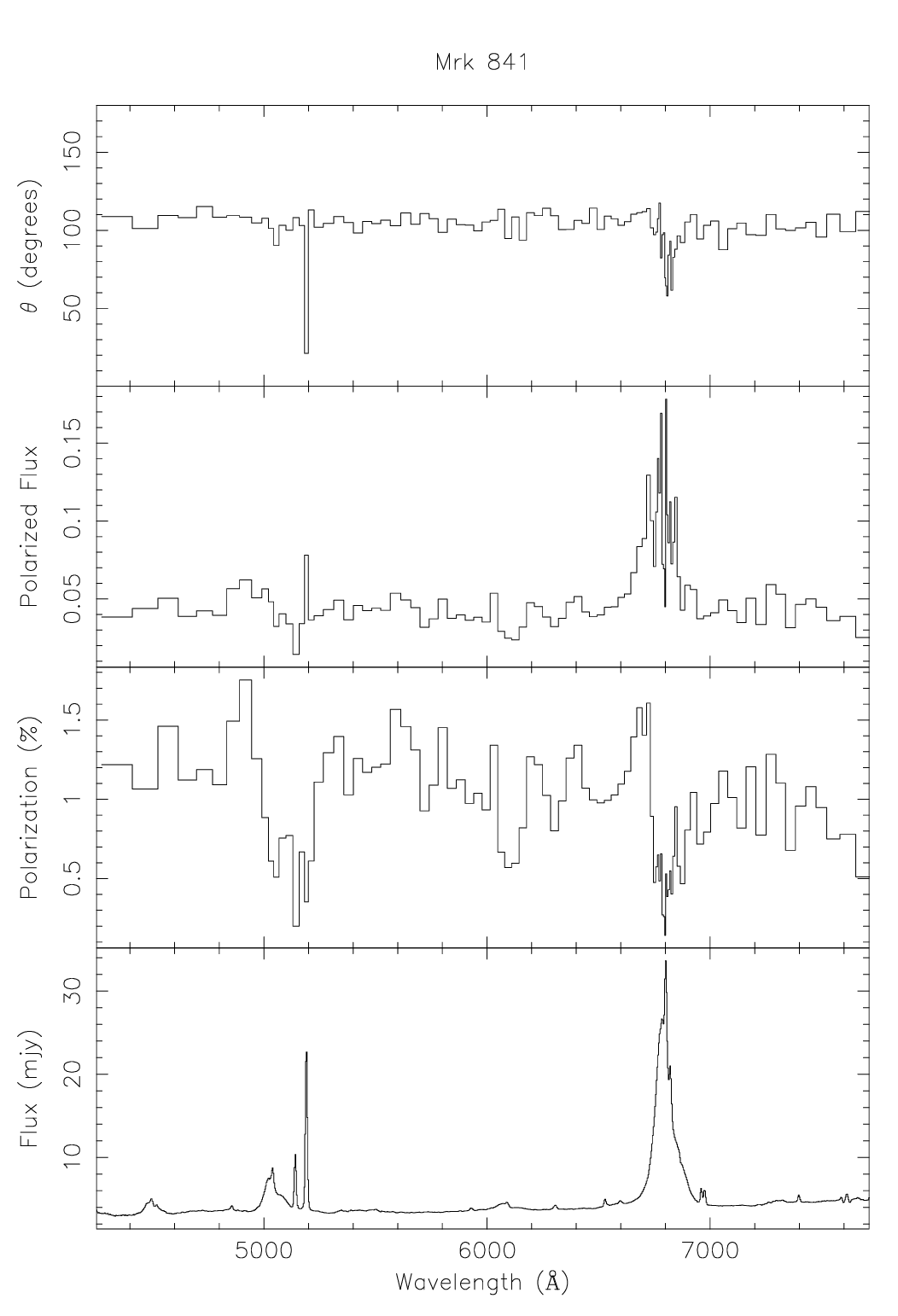,height=4.25in}

\centering{\caption{ As Fig. 1 for Mrk 841. Polarization data binned 
at 0.2 per cent.\label{fig:m841}}}

\end{figure}


\subsubsection{MRK\,279}
\label{m279}

Mrk\,279 is polarized at about 0.5 per cent, but shows a significant 
decrease in $p$ through the broad H$\alpha$ line (Fig.~\ref{fig:m279}).  
There is also a sharp change in $\theta$ over the line, which results 
in a large difference ($\sim 61\degr$) between the average values 
for the line and continuum.

There are no published radio maps of this source.  However, Ulvestad 
\& Wilson (1984a) report that a 6\,cm VLA map shows a resolved radio 
source with an axis in PA $\sim 90\degr$.  The average values of 
$\theta$ for the continuum and broad H$\alpha$ line both differ by 
$\sim 30^\circ$ from this PA but in opposite senses, the continuum and 
line having polarization PA's lower and higher than the radio PA 
respectively. In the continuum-subtracted polarization spectrum, 
$\theta$ is approximately constant over the H$\alpha$ line at 
$\sim 120^\circ$.

\subsubsection{MRK\,290}
\label{m290}

The continuum of Mrk\,290 is about 1 per cent polarized. The 
degree of polarization decreases strongly through the broad H$\alpha$ 
line but there is no significant change in $\theta$ across the 
spectrum (Fig.~\ref{fig:m290}).

M96 studied the optical polarization of Mrk\,290 in 1994 April.  He 
found similar continuum polarization (1.03 per cent in PA $\sim 150\degr$) and his data also show a decrease in $p$ associated 
with the H$\alpha$ line.  However, he measured an average broad 
H$\alpha$ polarization of only 0.10 per cent, rather lower than our 
value of 0.40$\pm$0.05 per cent.

The radio source of Mrk\,290 is unresolved in a 5\,GHz radio map 
obtained by Kellerman et al. (1994).

\begin{figure}

\psfig{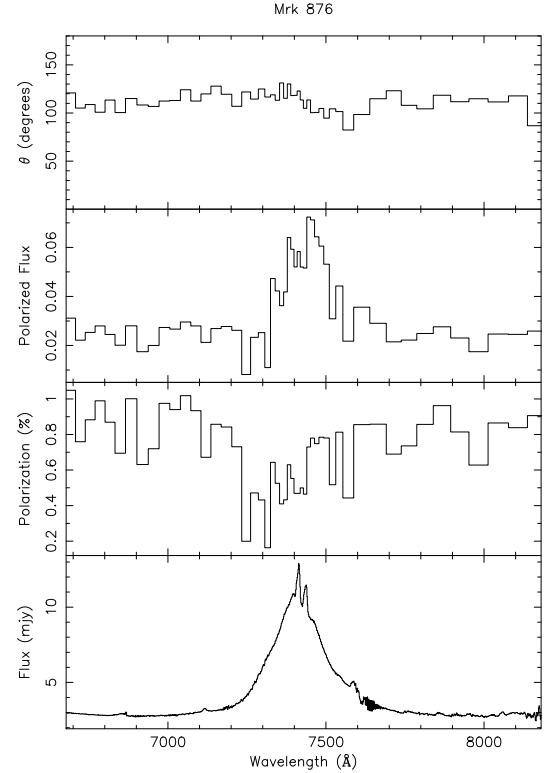}

\centering{\caption{As Fig. 1 for Mrk\,876. Polarization data binned 
at 0.15 per cent.\label{fig:m876}}}

\end{figure}


\begin{figure}

\psfig{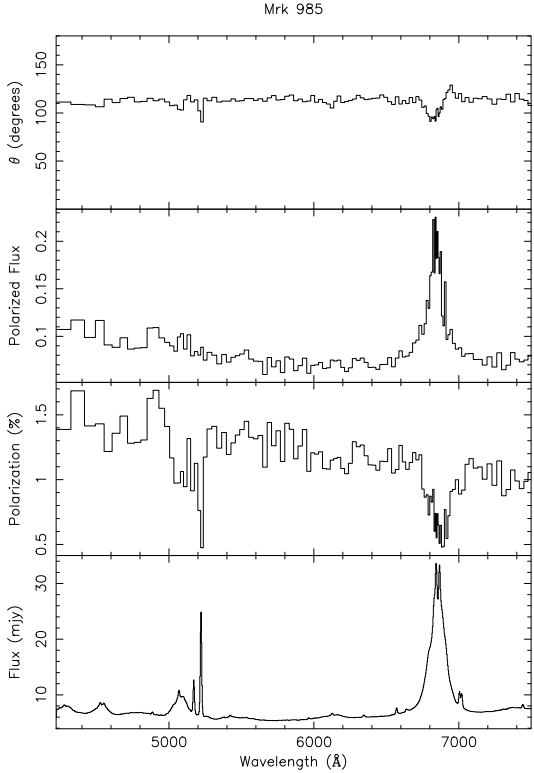}

\centering{\caption{ As Fig. 1 for Mrk\,985. Polarization data binned 
at 0.1 per cent.\label{fig:m985}}}

\end{figure}


\begin{figure}

\psfig{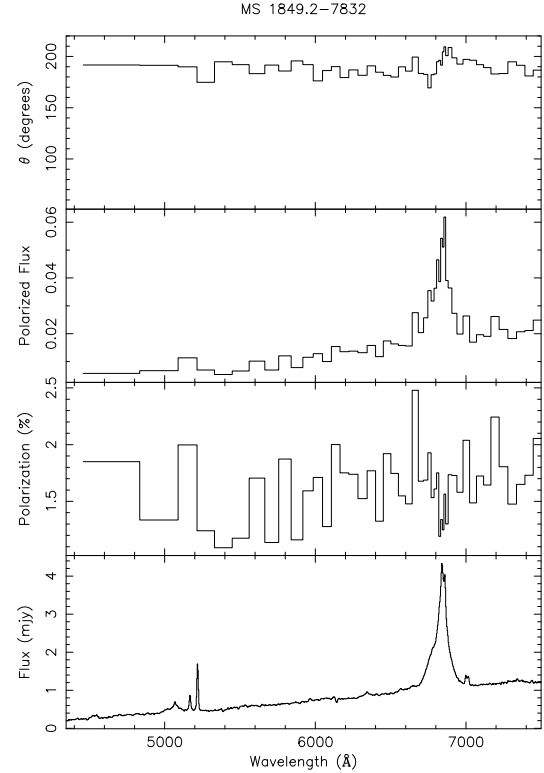}

\centering{\caption{ As Fig. 1 for MS\,1849.2$-$7832. Polarization 
data binned at 0.3 per 
cent.\label{fig:ms1849}}}

\end{figure}


\subsubsection{MRK\,304}
\label{m304}

The broad H$\alpha$ and H$\beta$ lines in Mrk\,304 both display 
prominent shoulders redwards of the line peak.

The percentage polarization spectrum shows structure across the broad 
H$\alpha$ line similar to that seen in Mrk\,6 and Mrk\,509 
(Fig.~\ref{fig:m304}).  The 
polarization increases from the continuum level ($\approx 0.5$ per cent) 
in the line wings but decreases to below that level through the line 
core.  The polarization error bins are wider at the blue end of the 
spectrum but there is a hint that similar structure is present in 
H$\beta$.  After continuum subtraction, the broad H$\alpha$ line has a 
higher average degree of polarization than the continuum.  This 
appears to be largely due to the increase in the line wings.  There is 
also a PA change of $\sim 30^\circ$ across the blue wing of the line.

M96 studied the optical polarization of Mrk 304 at various epochs.  
His August 1994 Keck observations show a continuum polarization of 
0.6 per cent at a PA of 132$^\circ$ and an average broad H$\alpha$ 
polarization of 0.62 per cent at a PA of 121$^\circ$.  Our 
measurements are consistent with these results.

M96 cites a private communication from Neff, reporting that in a 6\,cm 
VLA map Mrk\,304 has a double radio nucleus separated by 1.4$\arcsec$ 
(1.7 kpc) along a PA of 42$\degr$.  The average polarization PA's 
for both the continuum and broad H$\alpha$ are approximately 
orthogonal to the radio axis.

\subsubsection{MRK\,335}
\label{m335}

Mrk\,335 has been described as a `marginal' NLS1 (Giannuzzo et al. 
1998).

The continuum is weakly polarized at about 0.3 percent but $p$ 
increases significantly over the broad H$\alpha$ line 
(Fig.~\ref{fig:m335}).  The most 
prominent feature is a distinct peak in polarization ($p\sim 0.8$ per 
cent) in the blue wing of the line.  This feature appears to be 
associated with a sharp swing in $\theta$, by $\sim 80^\circ$ with 
respect to the value in the line core, which itself is polarized at a 
slighly smaller angle than the adjacent continuum.  Unusually for an 
object that exhibits a PA rotation, there is no decrease in 
polarization through the line core.

\subsubsection{MRK\,509}
\label{m509}

Our observations of Mrk\,509 have been discussed in detail by Young et 
al.  (1999) and are presented here for completeness.  The percentage 
polarization spectra exhibit structure similar to that seen in Mrk\,6 
and Mrk\,304.  At all 3 observation epochs $p$ increases above the 
continuum level in the broad H$\alpha$ line wings but decreases below 
that level at the line core (Fig.~\ref{fig:m509}).  A small rotation 
in $\theta$ is evident over the line profile.

Schmid et al. (2000) observed Mrk\,509 in 1999 August, 2 years after 
our last (1997 August) observation of this source. Their average 
polarization measurements are consistent with ours: for the H$\alpha$ 
line they find $p \approx 0.37$ per cent with $\theta \approx 149^\circ$ and for the 
continuum, $p \approx 0.64$ with $\theta \approx 152^\circ$. Their spectra also show 
similar variations in $p$ and $\theta$ across the H$\alpha$ profile.

Singh \& Westergaard (1992) present a 6\,cm VLA map showing 
elongations on opposite sides of the radio core along a PA of 
$\sim 165^\circ$.  These `jets' extend $\sim 1\arcsec$ (650 pc) from 
the radio core.  The average values of $\theta$ for the broad 
H$\alpha$ line and adjacent continuum differ by $\sim 20^\circ$ from 
the `jet' PA. At longer wavelengths (20\,cm) Singh \& Westergaard 
observe radio emission extending over $\sim 2\arcsec$ (1.3 kpc) along 
PA $\sim 125^\circ$.

\subsubsection{MRK\,841}
\label{m841}

This object also exhibits significant polarization structure across 
the broad emission-lines (Fig.~\ref{fig:m841}).  Broad minima in $p$ 
are associated with both the H$\alpha$ and H$\beta$ lines and there 
appears to be a similar dip corresponding to the He\,{\sc i} $\lambda 5876$ 
line.  In addition, both H$\alpha$ and H$\beta$ exhibit distinct peaks 
in $p$ associated with their blue wings at approximately the same 
velocity shift from the line peak.  However, there is no clear 
evidence for a corresponding polarization peak on the red side of the 
broad H$\alpha$ line, as is seen in Mrk\,509 and other sources.

The strong narrow lines (e.g., [O\,{\sc iii}] $\lambda\lambda 4959, 
5007$) also appear to depolarize the continuum.

\subsubsection{MRK\,876}
\label{m876}

Mrk\,876 is an extremely disturbed system with a double nucleus, tidal 
tails extending over $50\arcsec$ (85\,kpc) and a spiral companion 
(Hutchings \& Neff 1992).  The H$\alpha$ line is extremely broad; its 
wings extending beyond [S\,{\sc ii}] $\lambda$$\lambda$6716, 6731 in 
the red and [O\,{\sc i}] $\lambda$6300 in the blue.

There is a decrease in $p$ over the broad H$\alpha$ line, predominantly to 
the blue of the line peak.  There is also a swing in $\theta$ of 
$\sim30-40^\circ$ across the line, with the blue wing having a 
slightly larger, and the red a smaller, value than the adjoining 
continuum (Fig.~\ref{fig:m876}).

Kellerman et al.  (1994) present a 6\,cm VLA map showing that the 
radio source has faint extensions out to $\sim 1\arcsec$ (2.3 kpc) 
either side of the core in PA $\sim 140^\circ$.  The average 
polarization PA's for the continuum and broad H$\alpha$ line differ 
from the radio axis PA by $\sim 30^\circ$.

\subsubsection{MRK\,985}
\label{m985}

In the polarization spectrum both broad H$\alpha$ and H$\beta$ lines 
show decreases in $p$ relative to the continuum (Fig.~\ref{fig:m985}).  
Sharp dips in $p$ are also associated with the narrow [O\,{\sc iii}] 
lines.  There is a suggestion of an increase in polarization 
associated with the blue wing of H$\beta$ but no such feature is 
present in H$\alpha$.  In the continuum, there is a slow increase in 
$p$ towards the blue end of the spectrum.

There is a clear PA rotation of $\sim 40^\circ$ from red to blue 
across the broad H$\alpha$ profile, the blue and red wings having 
values of $\theta$ lower and higher than the adjoining continuum 
respectively.

\subsubsection{MS\,1849.2$-$7832}
\label{ms1849}

The data for this object are relatively noisy.  However, there appears 
to be a PA rotation on the red side of the broad H$\alpha$ line.  
There is also some evidence for a reduction in polarization over the 
core of the H$\alpha$ feature but this may be due to narrow-line 
region emission (Fig.~\ref{fig:ms1849}).

\begin{figure}

\psfig{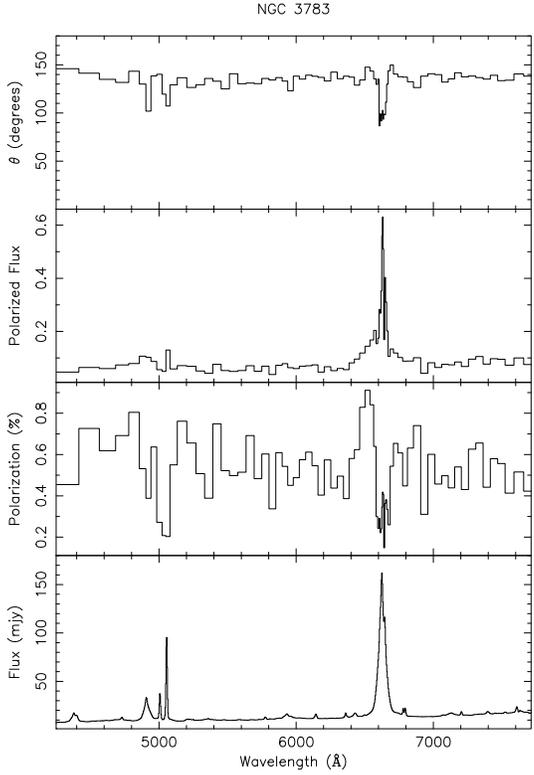}

\centering{\caption{As Fig. 1 for NGC\,3783. Polarization data binned 
at 0.1 per cent.\label{fig:n3783}}}

\end{figure}


\begin{figure}

\psfig{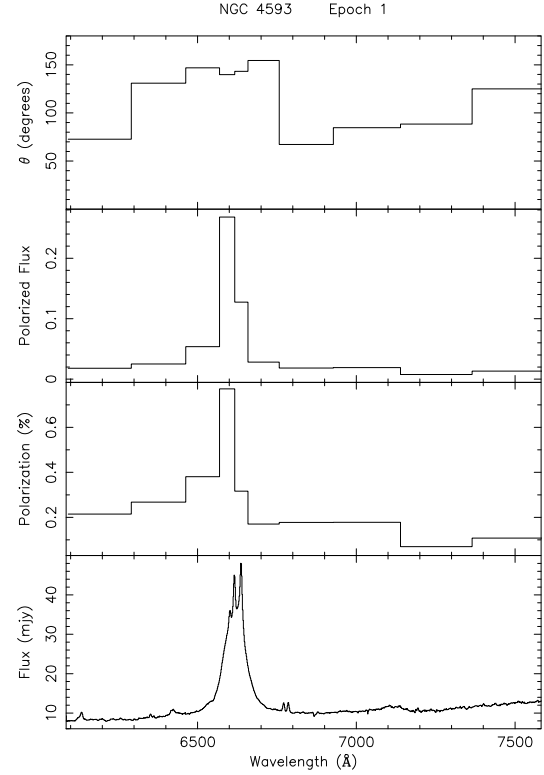}

\psfig{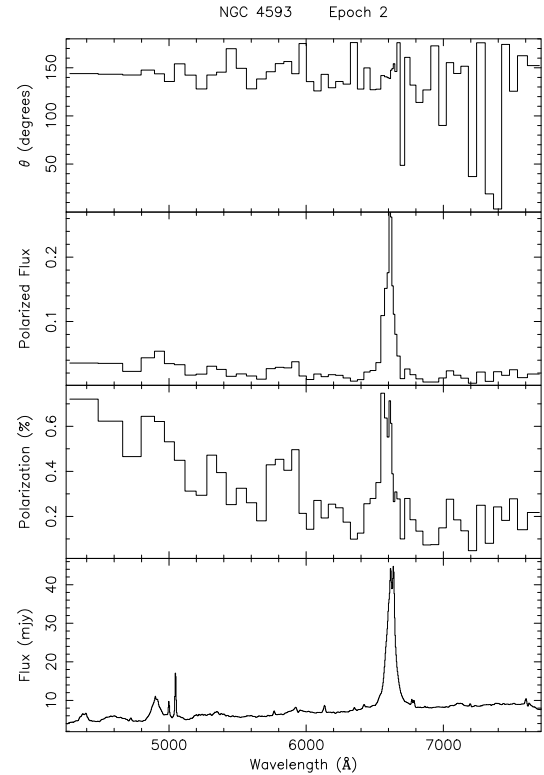}

\centering{\caption{ As Fig. 1 for NGC\,4593 -- 1997 February (Epoch 1),  
1997 May (Epoch 2). Polarization data binned at 0.1 per cent.\label{fig:n4593}}}

\end{figure}


\begin{figure}

\psfig{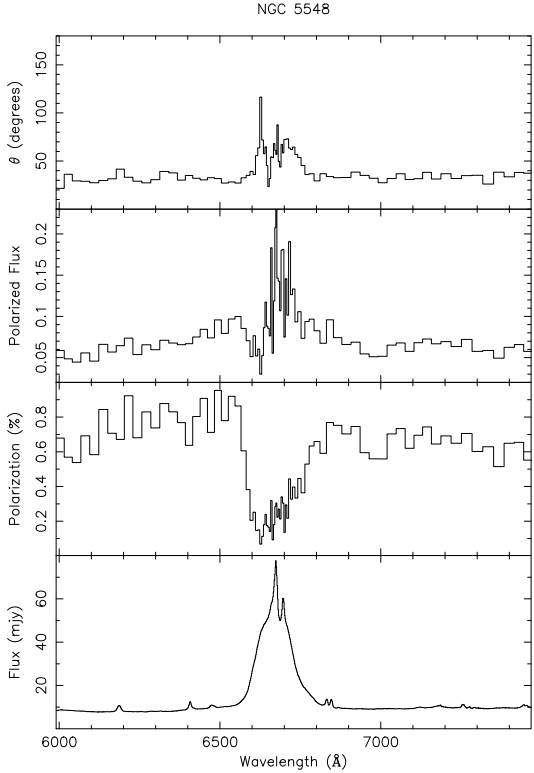}

\centering{\caption{ As Fig. 1 for NGC\,5548. Polarization data 
binned at 0.075 per cent.\label{fig:n5548}}}

\end{figure}


\begin{figure}

\psfig{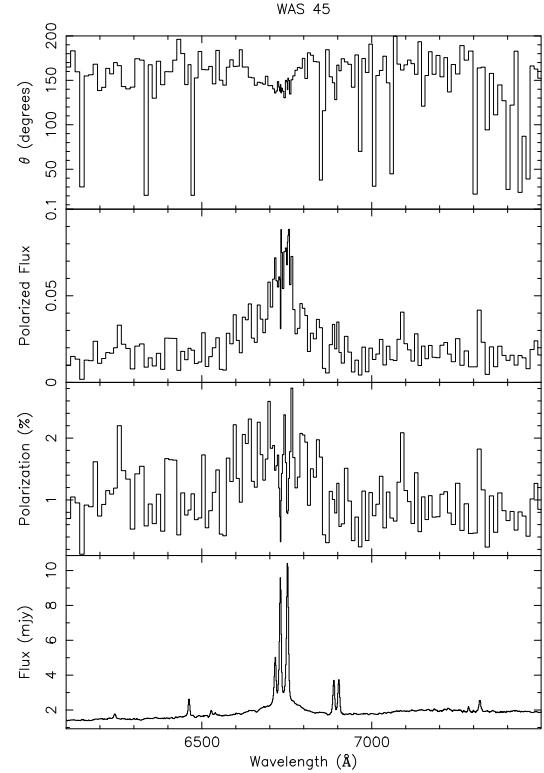}

\centering{\caption{ As Fig. 1 for Was\,45. Polarization data binned 
at 0.5 per cent.\label{fig:was45}}}

\end{figure}


\subsubsection{NGC\,3783}
\label{n3783}

The percentage polarization spectrum shows a peak associated with the 
blue wing of the broad H$\alpha$ line but $p$ decreases 
sharply through the core of the line.  There is also a dip in 
$p$ corresponding to the [O\,{\sc iii}] lines (Fig.~\ref{fig:n3783}).

The polarization PA changes sharply in the core of broad H$\alpha$ 
line (by $\sim 40\degr$) as a consequence of which, the average 
value of $\theta$ for the broad H$\alpha$ line differs from that of 
the continuum by $\sim 20\degr$.  The change in $\theta$ 
associated with the strong [O\,{\sc iii}] $\lambda 5007$ line suggest 
that the narrow-lines are polarized at a different PA to that of the 
continuum.  Some, but probably not all, of the PA change over H$\alpha$ 
may, therefore, be due to the blended narrow lines (the narrow 
component of H$\alpha$ and the [N\,{\sc ii}] $\lambda\lambda 6548, 6583$ 
doublet.  There is a suggestion that the wings of the broad H$\alpha$ 
line have a slightly larger value of $\theta$ than the continuum.

Although the spectrum is relatively noisy in the blue, the broad 
H$\beta$ line appears to have polarization properties similar to 
those of H$\alpha$.

\subsubsection{NGC\,4593}
\label{n4593}

We observed NGC\,4593 at two epochs separated by $\sim 110$ days 
(Fig.~\ref{fig:n4593}).  
The data obtained in our 1997 May observations cover a wider 
spectral range and are of much higher quality than our 1997 
February data.  The May data show that $p$ increases strongly to the 
blue in the continuum.  The broad H$\alpha$ line is much more 
highly polarized ($\sim 0.6$ per cent) than the adjacent continuum, 
for which $p\leq0.2$ per cent at the red end of the spectrum.  
The peak in $p$ associated with H$\alpha$ is blue-shifted relative 
to the line peak.  There is also an apparent increase in $p$ 
at around $5800\mathrm{\AA}$, which is perhaps associated with the 
broad He\,{\sc i} $\lambda$5876 line, but this feature is sensitive 
to the error binning and may, therefore, be spurious.

The blue-shifted polarization peak associated with H$\alpha$ is also 
clearly present in our poorer quality 1997 February data.  The average 
continuum polarization measured from these data is consistent with the 
second epoch.  It is not possible to compare the broad H$\alpha$ 
polarizations for the two epochs since we were unable to perform a 
satisfactory continuum subtraction for the earlier data.

\subsubsection{NGC\,5548}
\label{n5548}

Our data show that the broad H$\alpha$ line is polarized at a much 
lower level than the adjacent continuum (Fig.~\ref{fig:n5548}).  
Indeed, the degree of polarization in the broad H$\alpha$ line is one 
of the lowest that we were able to measure in our sample.  There is 
a sharp drop in $p$ on the blue side of the line and a gradual 
increase over the core and red wing back to the continuum level.  The 
sharp drop produces a corresponding dip in the polarized flux profile 
which, as a result, appears to show an isolated component in the blue 
wing.

The polarization PA spectrum displays unusual structure across the 
broad H$\alpha$ line.  There are two separate PA swings, each of about 
40$\degr$ relative to the continuum, in the blue and red sides of 
the line profile.  These excursions are separated by a narrow 
interregnum, just bluewards of the line peak, in which the value of 
$\theta$ is comparable to that of the continuum.  It seems unlikely 
that this is due to contamination by narrow-line emission polarized at 
a different PA, since it does not coincide in wavelength with the 
narrow component of the H$\alpha$ line, the strongest of the blended 
narrow-lines.

Data obtained by M96 show similar features, including the changes 
in $\theta$ and the separate blue-shifted structure in the polarized 
flux profile (as do data obtained by GM94).  M96 observed NGC 5548 at 
3 epochs and found the continuum near H$\alpha$ to be polarized at 
between 0.65 and 0.48 per cent whilst H$\alpha$ itself was polarized 
at between 0.17 and 0.24 per cent.  GM94 observed this object at 2 
epochs and found no variation in the average polarization, which they 
measured to be 0.67 per cent at a PA of 45.4$\degr$ over the 
wavelength range $4412 - 7179\mathrm{\AA}$.  Our polarization 
measurements are consistent with these earlier results.

Wilson and Ulvestad (1982) present 4885 and 1465\,MHz VLA maps showing 
lobes of radio emission extending $\sim 5\arcsec$ (1.6 kpc) to either 
side of an unresolved core in PA $\sim 155\degr$.  The average 
value of $\theta$ for the continuum differs by $\sim55\degr$ from 
the radio axis PA whereas the average polarization PA for the broad 
H$\alpha$ line is approximately orthogonal to the radio axis.

\subsubsection{Was\,45}
\label{was45}

In this Seyfert 1.8, the broad component of the H$\alpha$ line is 
relatively weak compared to the narrow-lines.  However, the broad 
component has a significantly higher polarization than the adjoining 
continuum ($p\sim 2$, compared with $\sim 1$ per cent) and as a 
result, is rather more prominent in the polarized flux spectrum than 
in the total flux spectrum (Fig.~\ref{fig:was45}). A small PA 
change is apparent across the broad H$\alpha$ line, the average values of 
$\theta$ in the line and continuum differing by $\sim 20\degr$.

Goodrich (1989a) made spectropolarimetric observations covering the 
wavelength range 4416 -- 7190$\mathrm{\AA}$ and measured an average 
polarization of $0.78\pm0.04$ per cent at $157.8\pm1.4\degr$.  He 
concluded that his measurements were entirely consistent with 
interstellar polarization.  However, the Galactic extinction in the 
direction of Was\,45 is E(B$-$V)$=0.018$, and therefore we would 
expect a maximum interstellar polarization of only 0.16 per cent.  
Since we detect degrees of polarization of 1 to 2 per cent, and also 
see significant changes in both the $p$ and $\theta$ in our data, we 
conclude that the measured polarization is largely intrinsic to the 
source.  In addition, inspection of the percentage polarization 
spectrum reveals dips in $p$ coincident with the narrow H$\alpha$ and 
[N\,{\sc ii}] $\lambda$6583 lines, conclusive evidence that at least 
some of the observed polarization is intrinsic to the source.

\subsection[]{Objects lacking clear evidence for intrinsic polarization}
\label{nopol}

The data for nine objects are too noisy for firm evidence of 
polarization intrinsic to the AGN to be discerned.  These objects 
(ESO\,113$-$IG45, ESO\,012$-$G21, Mrk\,705, Mrk\,915, Mrk\,926, 
NGC\,6104, NGC\,7213, PG\,1211+143 and UGC\,3478) may simply have 
levels of intrinsic polarization that are too low to be detectable 
with the signal-to-noise that we were able to achieve.  In some cases, 
a high starlight contribution from the host galaxy to the optical 
continuum may significantly dilute the intrinsic polarization.

There is weak evidence for intrinsic polarization in a further six 
objects -- Mrk\,871, Mrk\,896, NGC\,3516, NGC\,6814, NGC\,7469 and 
NGC\,7603.  However, in these cases, we cannot rule out a substantial 
contribution from interstellar polarization. The Galactic extinction 
in the lines-of-sight to these objects suggests that the interstellar 
polarization is comparable with what we have measured.  Nevertheless, 
Mrk\,871, Mrk\,896 and NGC\,7603 show marginal evidence for increases 
in the degree of polarization over the broad H$\alpha$ line, relative 
to the continuum level.  In addition, several objects exhibit 
polarization PA's either closely parallel to or approximately 
orthogonal to their radio axes.  This behaviour is characteristically 
exhibited by Seyfert nuclei, and while we cannot 
exclude mere coincidence, suggests that we may be detecting intrinsic 
polarization in some cases.

In NGC\,7603 the average polarization PA of the continuum and broad 
H$\alpha$ line ($\sim 130\degr$) is approximately orthogonal to the 
radio axis PA of $\sim 30\degr$ reported by Kukula et al. (1995) and 
Van der Hulst, Crane \& Keel (1981).  Goodrich (1989a) measured an 
average optical polarization of $0.42\pm0.03$ per cent in PA 
$128.9\pm1.8\degr$ and concluded that this was consistent with 
interstellar polarization.  We measure a much lower level of 
polarization than Goodrich (1989a), but a comparable PA.

NGC\,3516 has a continuum polarization PA of $30.1\pm8.0\degr$, 
which is approximately parallel to a `jet-like' radio feature.  Miyaji 
et al.  (1992) present a 20\,cm map showing the `jet' to consist of two 
components separated by $\sim 2\arcsec$ (350 pc) in PA 
$\sim10\degr$.  M96 studied the optical polarization of NGC\,3516 
and detected both a PA rotation and a variation in the degree of 
polarization over the broad H$\alpha$ line.  He measured an average 
continuum polarization of 0.65 per cent in PA 173$\degr$ and found 
the H$\alpha$ line to be polarized at 0.44 per cent in PA 
163$\degr$.  Our results are significantly different 
(Table~\ref{tab:res}).  In 
particular, we measure a much lower continuum polarization, 
$p\approx 0.1\pm 0.04$ per cent, which is comparable with the typical 
interstellar polarization expected for the line-of-sight E(B$-$V).  It 
is unclear why there is such a large difference between our 
polarization measurements and those of M96.

NGC\,6814 has a measured polarization $\sim 1.8$ per cent, among the 
highest in our sample.  The polarization PA is also approximately 
orthogonal to the East--West extension of the radio source (Ulvestad 
\& Wilson 1984b).  However, this galaxy suffers comparatively large 
line-of-sight extinction and the measured degree of polarization is 
comparable to the maximum expected for Galactic interstellar 
polarization.  Furthermore, neither $p$ nor $\theta$ vary 
significantly over the spectrum, consistent with interstellar 
polarization being the dominant component.  GM94 similarly concluded 
that the polarization measured in NGC\,6814 is entirely interstellar 
in origin.

NGC\,7469 is weakly polarized at a level consistent with the expected 
interstellar polarization.  However, in the 1998 October observations 
(Fig.~\ref{fig:n7469}), the red wing of the broad H$\alpha$ line exhibits a marked 
increase in $p$ above the continuum level, a feature not present at 
the previous 2 epochs.  In addition, the presence of sharp dips in $p$ 
at the wavelengths of the narrow lines indicates that at least some of 
the measured polarization is intrinsic to the AGN. NGC\,7469 has 
previously been studied by GM94 who found it to be essentially 
unpolarized at 0.02$\pm$0.02 per cent.

\begin{figure}

\psfig{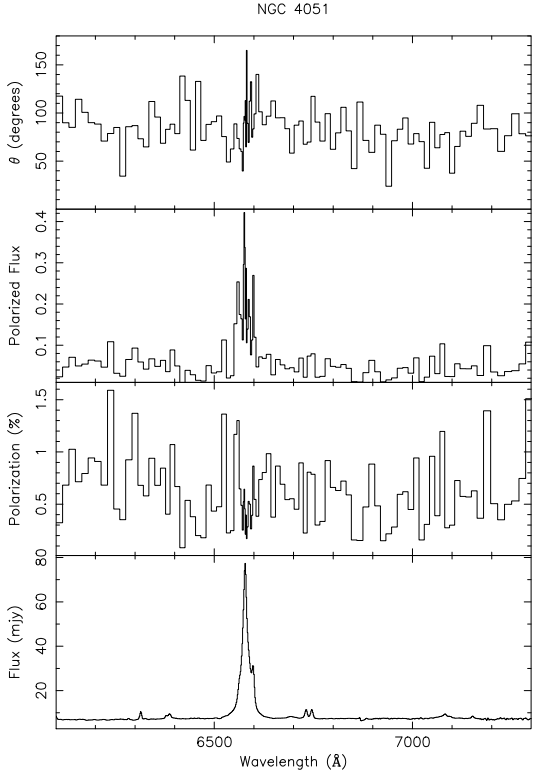}

\centering{\caption{ As Fig. 1 for NGC\,4051. Polarization data 
binned at 0.3 per cent.\label{fig:n4051}}}

\end{figure}

NGC\,4051 is included in this Section because its measured 
polarization does not vary with wavelength across the spectrum 
(Fig.~\ref{fig:n4051}). However, the line-of-sight Galactic extinction to NGC\,4051 is 
relatively small, implying a maximum interstellar polarization of 
0.12 per cent, which is significantly lower than the measured average
polarization ($\sim$ 0.5 per cent). Furthermore, the polarization PA 
is closely aligned with the radio source axis. Christopoulou et al. 
(1997) present a MERLIN 18\,cm map of NGC\,4051 showing a 
0.8$\arcsec$ (30 pc) triple radio source in PA $\sim 73\degr$, 
compared to a polarization PA of $\sim 80\degr$.

\begin{figure}
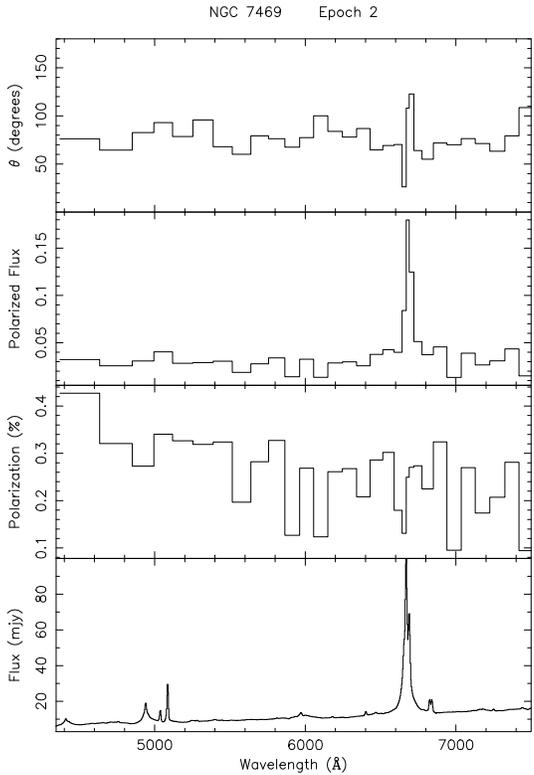


\psfig{figure=./n7469_e1.eps,height=4.25in}

\psfig{figure=./n7469_e2.eps,height=4.25in}

\centering{\caption{ As Fig. 1 for NGC\,7469 -- 1997 June (Epoch 1),  
1997 August (Epoch 2). Polarization data binned at 0.075 per cent.\label{fig:n7469}}}

\end{figure}


\begin{figure}

\psfig{figure=./n7469_e3.eps,height=4.25in}

\centering{\contcaption{(c) 1998 October.}}

\end{figure}


\section{Discussion}
\label{discussion}

We have detected intrinsic polarization in 20 Type 1 Seyfert nuclei.  
The polarization spectra of these sources all exhibit variations with 
wavelength associated with the broad H$\alpha$ line in either the 
degree or position angle of polarization, or both.  Some objects also 
exhibit wavelength-dependent continuum polarization.  In this Section, 
we discuss mechanisms by which rotations in position angle and 
changes in percentage polarization may arise.  We also outline a model 
involving a rotating emission disk and two scattering regions, which 
can at least partly account for the diversity in polarization 
properties exhibited by our sample.

\subsection{Rotations in polarization position angle}
\label{parot}

\begin{figure}

\psfig{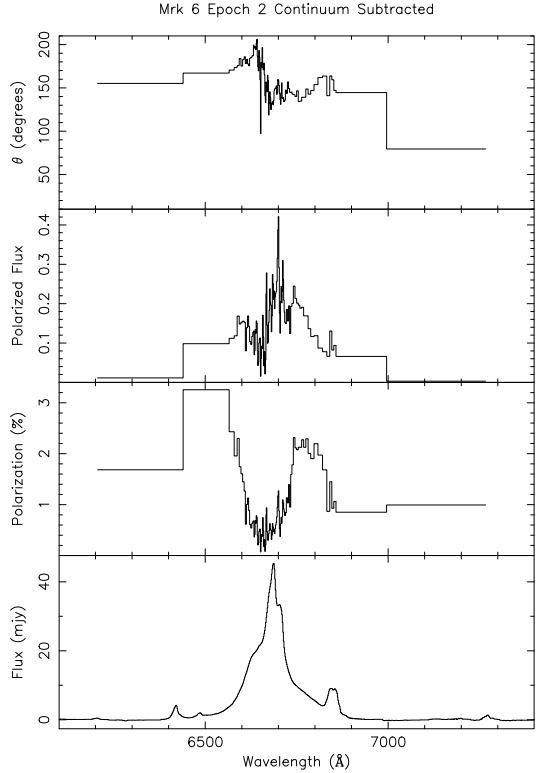}

\centering{\caption{continuum-subtracted polarization spectra of Mrk 
\,6 (1998 October data), showing the full amplitude of the PA rotation 
across the H$\alpha$ profile ($\sim 60\degr$, compared to $\sim 
40\degr$ in the unsubtracted spectrum shown in Fig.~\ref{fig:m6}).  Data 
binned at 0.3 per cent.\label{fig:m6cs}}}

\end{figure}

Our spectropolarimetric data show that rotations in polarization PA 
associated with the broad H$\alpha$ line are a common characteristic 
of Seyfert 1 nuclei; three-quarters of the intrinsically polarized 
objects exhibit significant rotations ranging up to $60-70\degr$.  
In some cases, this effect may arise from a 
difference in polarization PA between the H$\alpha$ and continuum 
emission.  However, in several sources (e.g., Akn\,120, 
KUV\,18217+6419, Mrk\,6, Mrk\,335 and Mrk\,985), the polarization PA 
changes as a function of line-of-sight velocity shift within the 
broad-line profile.  The full amplitude of these rotations in the 
H$\alpha$ polarization PA is only evident after subtracting the 
underlying continuum polarization.  In Mrk\,6, for example, there is 
an abrupt swing of $\sim 60\degr$ in the blue side of the profile 
core (Fig.~\ref{fig:m6cs}), whereas in Akn\,120 the H$\alpha$ polarization PA 
rotates monotonically by $\sim 70\degr$ across the profile from the 
blue to red wings (Fig.~\ref{fig:a120cs}).  These swings in polarization 
PA indicate that the orientation of scattering plane varies with 
line-of-sight velocity across the line profile which, in turn, places 
constraints on the geometry and kinematics of the scattering and/or 
emitting regions.

A velocity-dependent rotation in the broad emission line polarization 
PA requires either that the BLR is resolved by the scattering region or is 
subject to two distinct scattering regions that are arranged so that 
their scattering planes subtend a large angle to one another.  In 
either case, the velocity field in the BLR and/or scatterers must also 
be so ordered that there is some discrimination in wavelength for rays 
following different scattering paths.  Thus, if two scattering 
regions produce line profiles covering somewhat different wavelength 
ranges (because, one is shifted, or broadened, with respect to the 
other), the ratio of polarized fluxes from each scattering geometry 
will vary across the total line profile, resulting in a rotation in 
polarization PA. This could occur if the line profile produced by the 
BLR is orientation-dependent, as would be the case if, for example, 
the emitting gas resides in a rotating disk.  Scatterers located along 
the poles of the disk will `see' a narrower line profile than those in 
or near the equatorial plane.  The scattered line profiles from these 
two locations will therefore have different widths and, since the 
scattering planes are perpendicular, orthogonal polarizations.  The 
variation across the line profile in the amounts of polarized flux 
from the two scattering regions will produce corresponding changes 
in the polarization PA.

If, on the other hand, the BLR velocity field is isotropic (e.g., an 
ensemble of line-emitting clouds following virialized Keplerian 
orbits), motions in the scattering media are needed to provide the 
necessary discrimination in wavelength space.  Thus, an outflowing 
scattering region will tend to impart a red-shift to the scattered 
profile and the combination of outflowing polar and stationary 
equatorial\footnote{In this case, we may define the `equatorial' and 
`polar' scattering regions in relation to the circum-nuclear torus.} 
scattering regions would, in certain orientations, result in a 
rotation in the polarization PA across the line profile.

A rotation in polarization PA can also be produced if the scattering 
region is close enough to the BLR that the latter appears spatially 
extended as seen by the scatterers (near-field scattering).  Returning 
to the example of a rotating line-emitting disk, blue- and red-shifted 
emission from opposite sides of the disk will have different 
scattering angles at the same scattering element.  If the scatterers 
lie near the equatorial plane of the disk, this will, in general, 
result in the red and blue sides of the broad-line profile being 
polarized at different PA's.  Cohen et al.  (1999) have proposed that 
the PA rotation observed across the broad H$\alpha$ line in the radio 
galaxy 3C\,445 can be explained in this way and recently, Cohen \& 
Martel (2001) have discussed a simple scattering model based on this 
picture in relation to both 3C\,445 and the Seyfert 1 galaxy Mrk\,231.  
We will consider a similar model in relation to our own results in 
Section~\ref{model}.

\subsection[]{The radio axis--optical polarization relationship}
\label{radiopa}

It is well established that Type 1 Seyferts and broad-line radio 
galaxies tend to exhibit optical polarization PA's parallel to their 
radio axes whereas Type 2 Seyferts and narrow-line radio galaxies have 
optical polarizations orthogonal to the radio axis (e.g., Antonucci 
1983, 1984, 2001).  More recently, in his spectropolarimetric study of 
Seyfert 1s, Martel (M96) finds a preference for parallel alignments 
but also shows that when the small-scale radio axis is considered, 
there is an apparently bi-modal distribution, with several objects 
having an optical polarization PA orthogonal to this axis.

We have been able to obtain information on the radio axis for 10 of 
the objects in which we have detected polarization intrinsic to the 
AGN. The differences ($\Delta\mathrm{PA}$) between the PA of the 
projected axis of the radio source and the average polarization PA's 
of, respectively, the broad H$\alpha$ line and the adjacent continuum, 
are listed for these objects in Table~\ref{tab:dpa}.  The radio axis PA's are taken 
from the references cited in Section~\ref{results} and are listed in the table.  We 
have either used the published value, or where none is available, made 
our own estimate from a published radio map.  In Seyfert galaxies, it 
is often difficult to establish the radio source axis unambiguously.  
We have used the PA of obvious `jet-like' features if one is present, 
otherwise we have used the PA of the innermost contour tracing an 
elongation in the source.

Considering the H$\alpha$ polarization PA, 6 of the objects listed in 
Table ~\ref{tab:dpa} have $\Delta\mathrm{PA}\le 30^\circ$, whilst 3 have 
$\Delta\mathrm{PA}\ge 70^\circ$.  In only one object, 
KUV\,18217+6419, is the PA difference intermediate between these two 
ranges.  The values of $\Delta\mathrm{PA}$ for the continuum are 
similarly distributed.  Like M96, therefore, we find a dichotomy in 
the alignment of the optical polarization {\bf E} vectors (both 
H$\alpha$ and the adjacent continuum) with the radio source axis, the 
{\bf E} vectors being roughly parallel to this axis in the majority of 
objects but roughly perpendicular to it in a significant minority.  
Note, however, that 4 of the objects listed in Table ~\ref{tab:dpa} also appear in 
M96's sample, two having parallel (Akn\,120 and Mrk\,509) and two 
perpendicular (Mrk\,304 and NGC\,5548) alignments.

\begin{table*}

\centering


\caption{Differences between the average polarization PA's of the 
broad H$\alpha$ line and the adjacent continuum and the PA of the 
radio source. PA's are given in the range 0--180$^\circ$ (e.g., 
320$^\circ$ = 140$^\circ$).}

\begin{tabular}{@{}rcccccl@{}}

\hline

   Object&Epoch	&   Radio PA	& Scale	&$\Delta\mathrm{PA}$ (H$\alpha$) 
&$\Delta\mathrm{PA}$ (Cont.) & Notes \\

         &      &  ($^\circ$)   & (kpc) &  ($^\circ$)      	&  
($^\circ$) 	 &       \\

\hline

Akn 120	 &Feb 97&    50         & 6.10  &  15   &  21   & Contour 
elongations\\

	 &Oct 98&               &       &  25	&  27   &                    \\

Akn 564	 &	&   170		& 0.46  &  78   &  83	& Jet--like\\

IZw1	 &Aug 97&   140		& 34.4  &  7	&  6	& Contour elongations\\

         &Oct 98&		&	&  6	&  12	&  \\

KUV 18217+6419& &   20		& 0.54	&  44	&  57	& Jet--like\\

Mrk 6	 &Oct 98&   170		& 0.14	&15	&  14	& Jet--like\\  

Mrk 279	 &	&	90	&   ?	&30	&  30	& Unpublished radio map\\

Mrk 304	 &	&	42	& 1.70  &81	&  86	& Unpublished radio map\\

Mrk 509	 &Aug 96&	165	& 0.65	&20	&  13	&Jet--like\\

   	 &May 97&		&	&17	&  24	&\\

	 &Aug 97&		&	&20	&  26	&\\

Mrk 876	 &	&	140	& 2.30	&28	&  30	&Contour elongations\\

NGC 5548 &	&	155	& 1.63  &88	&  58	&Jet--like\\

\hline

\end{tabular}

\label{tab:dpa}
\end{table*}

The simplest interpretation is that the polarized light in the 
`parallel' and `perpendicular' groups is produced by different 
scattering geometries.  Scattering in a plane perpendicular to the 
radio jet axis would tend to produce polarization parallel to that 
axis, whereas scattering in a plane aligned with the jet axis would 
produce orthogonal polarization.  Assuming that the radio jet is 
co-aligned with the torus axis, parallel and perpendicular 
polarizations correspond to scattering in the equatorial plane of the 
torus and along its poles, respectively.

However, as we have noted, several objects exhibit large swings in 
polarization PA across their broad H$\alpha$ lines.  This suggests 
that considering only the {\em average} polarization PA for the line 
is too simplistic and may cause us to miss important clues to the 
scattering geometry.  Therefore, for each of the galaxies in Table~\ref{tab:dpa}, 
we have used the continuum-subtracted spectra to calculate the 
polarization PA's in three separate wavelength bins corresponding to 
the red wing, core and blue wing of the broad H$\alpha$ line profile.  
These polarization PA's are listed, together with the radio axis PA, in 
Table~\ref{tab:hapa}.

In the three objects of the `perpendicular' group (Akn\,564, Mrk\,304 
and NGC\,5548) large PA differences with respect to the radio axis are 
maintained across the line profile, although with some variation 
between bins.  Other objects exhibit more complex relationships 
between the polarization {\bf E} vector and the radio axis.  At first 
sight, KUV\,18217+6419 is anomalous in that its average broad 
H$\alpha$ polarization PA differs by $44\degr$ from that of its 
`jet-like' radio source and therefore falls into neither `parallel' 
nor `perpendicular' groups.  However, the binned PA measurements show 
a rotation such that the {\bf E} vector is roughly perpendicular to 
the radio axis in the blue wing ($\Delta \mathrm{PA} \sim 72\degr$) 
but nearly parallel in the red wing ($\Delta \mathrm{PA} \sim 
9\degr$).

\begin{figure}
\psfig{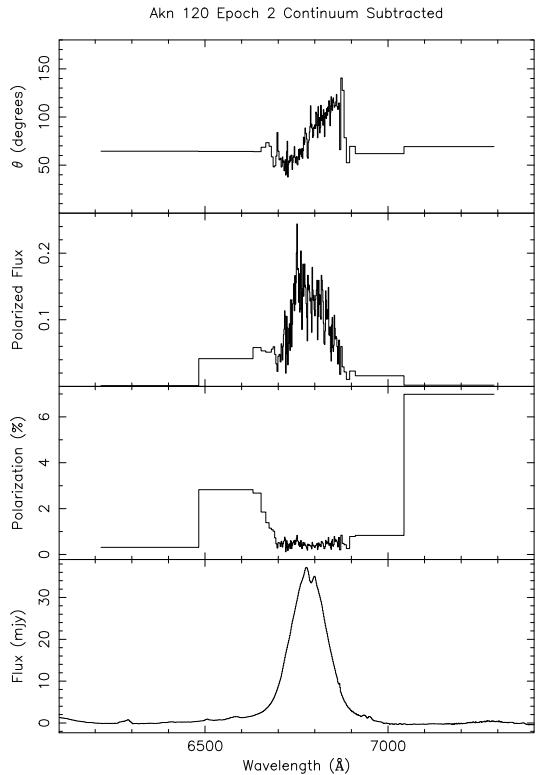}
\centering{\caption{Continuum-subtracted spectra of Akn 120 from 1998 
October, showing the large rotation in $\theta$ across the line 
profile. Polarization data binned at 0.3 per cent.\label{fig:a120cs}}}
\end{figure}

Akn\,120 falls comfortably into the `parallel' group in Table~\ref{tab:dpa}, but 
there is a similar change in {\bf E} vector alignment across the broad H$\alpha$ 
profile.  This is not very apparent from the binned PA measurements 
listed in Table~\ref{tab:hapa}, but inspection of the continuum-subtracted spectra 
from our 1998 October observations (Fig.~\ref{fig:a120cs}) reveals a PA swing from 
$\sim 50\degr$ in the blue wing to $\sim 120-130\degr$ in the 
red.  So while, on average, the broad H$\alpha$ PA differs by 
25$\degr$ from that of the radio axis, the {\bf E} vector rotates 
from a parallel alignment ($\Delta \mathrm{PA} \sim 0\degr$) in the 
blue wing to a roughly perpendicular orientation in the red wing 
($\Delta \mathrm{PA} \sim 70-80\degr$).

In at least three other `parallel' objects, the blue and red wing 
polarization PA's straddle the radio source PA, the {\bf E} vectors, 
typically pointing a few 10's of degrees to either side of the 
projected radio axis.  As discussed in Section~\ref{model}, this behaviour is 
consistent with near-field scattering of line emission from a 
rotating disk, by scatterers co-planar with the disk.

In making these comparisons, it is important to bear in mind the 
uncertainties inherent in establishing the radio axis PA. One problem, 
already alluded to, is that the radio axis may not be well-defined in 
partially resolved or amorphous sources.  Another is that, even when 
detailed maps are available, the large scale ($\sim$ kpc) structure of 
the radio source often has a different axis to that of the core.  In 
these cases, it is not clear whether the systemic axis of the AGN is 
best represented by the large or small scale structure.  For example, 
if the radio jet is precessing, the large scale axis may represent the 
time-averaged direction of the radio jet, in which case, it would be 
the most appropriate axis with which to compare the polarization PA. 
On the other hand, the large scale radio structure may itself be 
significantly influenced by the circumnuclear environment.  For 
example, once the radio ejecta has spent its kinetic energy, the 
plasma will tend to expand along the local pressure gradient and 
might, therefore, become more elongated along the galactic minor axis 
than the jet axis.  It is also possible that kiloparsec-scale radio 
lobes are contaminated or even dominated by starbursts (e.g., Baum et 
al.  1993).

\begin{table*}

\centering


\caption{Radio axes and the polarization PA's in the core and wings 
of the broad H$\alpha$ line.}

\begin{tabular}{@{}rccccccc@{}}

\hline

   Object&Epoch	&   Radio PA	& Scale	&Blue Wing PA&Core PA   &Red 
Wing PA&  Average broad H$\alpha$ PA \\

         &      & ($^\circ$)    & (kpc) &($^\circ$)  
&($^\circ$)&($^\circ$) &      ($^\circ$)         \\

\hline

Akn 120	 &Feb 97&    50         & 6.10  &65.0$\pm$2.2  &74.0$\pm$2.6 
&67.1$\pm$2.3 &  64.9$\pm$5.3\\

	 &Oct 98&               &       &62.1$\pm$1.7  &79.0$\pm$1.4 
&91.9$\pm$7.2 &  74.9$\pm$1.2\\

Akn 564	 &	&   170		& 0.46  &113.8$\pm$14.6&92.8$\pm$4.0 
&73.8$\pm$9.4  & 92.5$\pm$3.9\\

IZw1	 &Aug 97&   140		& 34.4  
&119.9$\pm$12.9&126.0$\pm$8.1&166.3$\pm$16.0 & 132.6$\pm$7.6\\

         &Oct 98&		&	&121.7$\pm$5.1 &129.3$\pm$2.3&151.5$\pm$3.2  &  
133.8$\pm$1.9\\

KUV 18217+6419& &   20		& 0.54	&127.9$\pm$4.1 
&170.1$\pm$6.7&11.1$\pm$7.5  & 156.5$\pm$5.5\\

Mrk 6    &Oct 98&   170		& 0.14	&175.0$\pm$1.3 &148.2$\pm$1.8&147.3$\pm$1.1  
&155.2$\pm$0.9\\

Mrk 279	 &	&	90	&   ?	&79.2$\pm$18.8&133.9$\pm$7.3&109.3$\pm$11.5 
&120.3$\pm$7.8\\

Mrk 304  &	&	42	& 1.70  &122.5$\pm$3.3&111.3$\pm$5.6&139.5$\pm$6.6  
&123.3$\pm$2.9\\

Mrk 509	 &Aug 96&	165	& 0.65	
&152.7$\pm$2.1&135.8$\pm$2.7&147.9$\pm$3.1   &145.3$\pm$1.6\\

   	 &May 97&		&	&162.1$\pm$3.3&135.3$\pm$4.4&147.0$\pm$3.8   
&148.2$\pm$2.4\\

	 &Aug 97&		&	&156.8$\pm$4.5&130.9$\pm$4.4&147.5$\pm$3.1   
&145.0$\pm$2.5\\

Mrk 876	 &	&	140	& 2.30	&157.1$\pm$9.5&113.8$\pm$3.9&88.0$\pm$8.2   
&112.4$\pm$4.6\\

NGC 5548 &	&	155	& 1.63  &32.1$\pm$6.4&78.1$\pm$2.3&53.3$\pm$4.4    
&67.3$\pm$2.4\\

\hline

\end{tabular}

\label{tab:hapa}
\end{table*}

Nevertheless, if we take the radio axis PA's we have compiled at face 
value, there is evidence that two different scattering geometries, 
producing orthogonal polarizations, are present in Type 1 Seyfert 
nuclei.  The `parallel' group is presumably dominated by scattering in 
the equatorial plane of the torus and the `perpendicular' group by 
polar scattering along the radio jet/torus axis.  The very large rotations 
in polarization PA that occur across the broad H$\alpha$ line in 
KUV\,18217+6419 and Akn\,120, where the polarization changes from 
parallel to orthogonal to the radio axis, might imply that both scattering 
routes are important in some objects.

It is worth recalling that optical polarization perpendicular to the 
radio axis is a well-established characteristic of Type 2 Seyfert 
nuclei.  The presence of several perpendicularly polarized objects in 
our sample (Akn\,564, Mrk\,304 and NGC\,5548 from Table~\ref{tab:dpa} and possibly 
also NGC\,6814 and NGC\,7603 which, although they may be heavily 
contaminated by interstellar polarization, also have optical 
polarization PA's approximately orthogonal to the radio axis) provides 
evidence for a population of Seyfert 1s which, like Seyfert 2s, are 
dominated by polar scattering. However, Akn\,564, NGC\,5548 and, in particular, Mrk\,304, all display variations in $p$ over the broad-line, most notably depolarizations over the line core, that are more characteristic of equatorial, rather than polar, scattering as discussed in the following Sections.

\subsection[]{Polarization structure across the broad H$\alpha$ line}
\label{polstr}

Approximately three-quarters of the objects in which we have detected 
intrinsic polarization show significant decreases in polarization over 
the core of the broad H$\alpha$ line.  Cohen et al.  (1999) have 
proposed that similar features in broad-line radio galaxies are due to 
dilution of the polarized underlying continuum by unpolarized broad 
H$\alpha$ emission, the BLR being located outside the scattering 
regions.  However, simple dilution is not, in general, a satisfactory 
explanation for the polarization structure observed in our Seyfert 1 
sample.  The variations in $p$ do not mirror the shape of the 
broad-line profile and in particular, it is often the case that one or 
both line wings are more highly polarized than the continuum (e.g., Mrk\,6, 
Mrk\,304, Mrk\,509).  Weakly polarized narrow H$\alpha$ and [N\,{\sc 
ii}] $\lambda\lambda 6548, 6583$ lines may cause some dilution, but 
this cannot account for the large widths of the percentage 
polarization minima.

Alternatively, M96 has proposed a model in which variations in 
polarization across the broad H$\alpha$ line are ``produced by 
scattering of radiation emitted by the central continuum source from 
sub-regions of different physical properties in the BLR''.  However, 
the ad hoc nature of this picture makes it unattractive, and we prefer 
to look for interpretations that are consistent with the basic 
geometry imposed by the Seyfert unification scheme.

We note firstly, that polarization minima within the broad H$\alpha$ 
line do not necessarily indicate that the scatterers are interior to, 
or within, the BLR. For example, cancellation between two sources of 
orthogonally polarized flux overlapping in wavelength space could 
reduce the net polarization.  With the possible exception of Mrk\,279, 
the broad H$\alpha$ emission and underlying continuum do not exhibit 
orthogonal polarizations.  Therefore, if the decreases in $p$ in the 
line core are due to cancellation, then it must be between different 
sources of polarized broad H$\alpha$ emission.  Many of the objects 
that show polarization minima also exhibit PA rotations that might 
indicate the presence of more than one source of polarized H$\alpha$ 
emission.  However, the PA rotations are usually rather smaller than 
might be expected for orthogonally polarized sources and furthermore, 
polarization minima are also observed in objects which show no 
evidence for a PA rotation across the broad H$\alpha$ line (e.g.,  
Mrk\,290).

A more satisfactory explanation for the decreases in $p$ in the core 
of the broad H$\alpha$ line is that the line profile in total flux 
(the direct view) is narrower than that in polarized flux (scattered 
view).  This will occur naturally if the BLR is confined to a rotating 
disk and is surrounded by a co-planar ring of scatterers.  The 
scatterers view the disk edge-on whereas the direct line-of-sight to 
the BLR has to pass within the opening angle of the circum-nuclear 
torus and therefore provides a more face-on view (assuming that the 
disk and torus are roughly co-axial).  Thus, the line profile in 
scattered (polarized) light will always be broader than that in direct 
(unpolarized) light and the latter will dilute the polarization of the 
scattered component in the core of the combined line profile.  If the 
underlying continuum has a constant polarization, $p$ will increase 
relative to the continuum level in the line wings, but will drop below 
that level in the core.  This type of H$\alpha$ polarization structure 
-- a decrease in polarization over the line core flanked by increases 
in both line wings -- is clearly observed in 3 objects, Mrk\,6, 
Mrk\,304 and Mrk\,509.  A similar effect will, in general, result from 
electron scattering regardless of the exact scattering geometry, since 
the scattered line profile can be significantly broadened by the 
thermal velocity dispersion of the scattering electrons (e.g., the 
velocity dispersion for electrons at a temperature $T\sim 10^{6}$\,K 
is $\sim 5000$\,km\,s$^{-1}$).

A few objects, notably Fairall\,51, NGC\,4593 and Was\,45, show 
significant increases in polarization, relative to the continuum 
level, covering not just the wings, but the entire profile of the 
broad H$\alpha$ line -- including the core.  Similar broad peaks in 
polarization associated with the Balmer lines are present in those 
Type 2 Seyfert nuclei that exhibit Type 1 spectra in polarized flux.  
In these objects, the nuclear continuum source and BLR are hidden by 
the torus and only visible in scattered, hence polarized, light.  
However, since the continuum and line emission are presumably subject 
to the same scattering geometry (polar scattering) they should have 
the same polarization level.  The increase in $p$ over the broad 
H$\alpha$ line is explained by the presence of a further, unpolarized 
continuum component (likely to be starlight from the host galaxy) that 
dilutes the polarized light from both the nuclear continuum and the 
BLR. The contribution of polarized flux from the BLR lessens the 
effect of dilution over the broad-line profile, leading to an increase 
in percentage polarization.

In the Type 1 Seyferts, we have a direct view of the nuclear continuum 
source and BLR and the dilution effect of starlight would be small if 
it is dominated by the AGN continuum.  However, it seems plausible 
that, at least insofar as their scattering and polarization properties 
are concerned, Fairall\,51, NGC\,4593 and Was\,45, are more akin to 
Type 2 Seyferts than Type 1's.  In this respect, it is significant 
that in both Fairall\,51 and NGC\,4593 the continuum polarization 
increases to shorter wavelengths.\footnote{Unfortunately, we cannot 
ascertain the polarization wavelength dependence in Was\,45 since our 
data for this object only cover the H$\alpha$ region.} This is another 
property common to Type 2 Seyferts, where it is generally believed to 
indicate dust scattering of light from the obscured AGN continuum 
source combined with an unpolarized starlight contribution (or, 
alternatively, electron scattering combined with a red unpolarized 
continuum).  These 3 objects, again like Seyfert 2s, exhibit no 
significant PA rotation across the broad H$\alpha$ line.  Their 
optical polarization properties, therefore, are consistent with 
far-field polar scattering of the BLR and AGN continuum, combined with 
a relatively large (for Type 1 Seyferts) starlight fraction 
contributing to the unpolarized continuum flux.  The latter situation 
could arise, perhaps, from partial extinction of the AGN (which would 
also serve to reduce the polarized flux from near-field scattering 
regions which might otherwise produce PA rotations).  Schmid et al.  
(2001) reach similar conclusions from their detailed 
spectropolarimetric study of Fairall\,51.  After subtracting a 
starlight fraction estimated to be $\approx 40$ per cent (at $\approx 
8500$\AA), they find that the AGN spectrum can be explained by a 
combination of reddened direct light and dust-scattered polarized 
light. They suggest that in this galaxy, the direct line-of-sight to the 
AGN passes through the top layers of the dusty circum-nuclear torus, 
above which is a dusty polar scattering region.

A clear prediction of the polar scattering scenario is that the 
polarization {\bf E} vector should be orthogonal to the projected 
radio axis, as is the case in Type 2 Seyferts.  Unfortunately, we have 
been unable to find published radio maps for Fairall\,51, NGC\,4593 and 
Was 45 with which to verify this prediction.  Nevertheless, in 
Mrk\,704, another Type 1 Seyfert (not included in our sample) which 
has very similar polarization properties, the optical polarization PA 
{\em is} known to be perpendicular the radio axis (GM94; M96).  There 
is, therefore, good evidence that the polarization properties of a 
significant number of Seyfert 1s (roughly 10 per cent of our total 
sample) are dominated by Seyfert 2-like polar scattering.

However, other objects exhibiting somewhat similar polarization properties 
(i.e., an increase in polarization over the entire broad H$\alpha$ 
line and/or polarization PA orthogonal to the radio axis) cannot be 
explained in terms of simple polar scattering.  For example, Mrk\,335 
exhibits a large PA rotation across broad H$\alpha$ which is not 
generally seen in Seyfert 2s.  Akn\,564 and NGC\,5548 both have 
polarization PA's orthogonal to their radio source axes, but exhibit 
decreases in polarization over the core of the 
line. In addition, whilst Mrk 304 shows a depolarization over the broad 
H$\alpha$ line core flanked by increases in both line wings, consistent 
with scattering from the equatorial plane of a rotating emission disk, 
it has a optical polarization PA $\em{orthogonal}$ to the reported radio 
source PA. However, it is still possible to understand the apparently 
conflicting behaviours of $\theta$ and $p$ in terms of equatorial scattering, 
as discussed in Section~\ref{eqscat}.

\subsection{Other considerations}
\label{other}

\subsubsection{The polarization of narrow-line Seyfert 1 galaxies}
\label{nls1}

Narrow-line Seyfert 1 galaxies (NLS1; Osterbrock \& Pogge 1985) have 
aroused much interest in recent years as it has become increasingly 
clear that they exhibit extreme optical and X-ray properties (e.g., 
Grupe 2000).  As a sub-class, however, they do not appear to exhibit 
unusual optical polarization; most are weakly polarized although a few 
exhibit wavelength-dependent polarization at levels $>2$ per cent 
(Goodrich 1989b; Grupe et al.  1998).  Nevertheless, among 
various explanations for the narrowness of the Balmer lines, it has 
been proposed (i) that the BLR is partly obscured and (ii) that the 
Balmer lines are emitted by a disk, which, in NLS1's, happens to have 
a face-on orientation.  Since both scenarios might be expected to 
produce certain polarization signatures, it is worth examining the 
Balmer-line polarization properties of the NLS1's in our sample.

There are 8 objects in our sample that have been classified as NLS1's.  
We have detected intrinsic polarization in only three: Akn\,564, 
I\,Zw1 and Mrk\,335.  The first two both exhibit decreases in 
polarization across the H$\alpha$ line, but differ in the orientation 
of the polarization {\bf E} vector relative to the radio axis.  In 
Akn\,564, the average line and continuum polarization PA's are 
approximately orthogonal to the radio axis.  On the other hand, I\,Zw1 
is, on average, polarized parallel to its radio axis and it is also 
one of the objects in which there is a PA swing across the broad 
H$\alpha$ line such that the blue and red wing polarization PA's 
bracket that of the radio axis (Table~\ref{tab:hapa}).  The other NLS1 for which we have obtained a radio axis is NGC\,4051.  We cannot be certain that 
the measured 0.5 per cent polarization is intrinsic to the AGN in this 
source (Section~\ref{nopol}), but if so, the polarization PA is approximately 
parallel to the radio axis.

The classification of Mrk\,335 as a NLS1 is somewhat dubious 
(Giannuzzo et al.  1998), but we note nevertheless, that unlike 
Akn\,564 and I\,Zw1, the degree of polarization increases over the 
broad H$\alpha$ line.  Mrk\,335 is also unusual in that its blue wing 
is polarized at a higher level and at a roughly orthogonal PA compared 
to the rest of the line and the local continuum.

With the exception of Mrk\,335 (and possibly I\,Zw1), there is no 
evidence that, in the NLS1, the H$\alpha$ line wings are broader in 
polarized than in total flux. This argues against obscuration of 
the inner regions of the BLR as the explanation for the relatively 
narrow broad-line profiles.  The observations similarly do not really 
support the idea that the face-on orientation of a disk-like BLR is 
the main characteristic that differentiates NLS1s from `normal' Seyfert 1s.  
In this case, we would expect them to exhibit similar H$\alpha$ 
polarization properties (in the context of the orientation sequence 
outlined in Section~\ref{eq+pol}, NLS1's would be weakly polarized).  In 
fact, as we have already noted, there are large differences.  
Collectively, the H$\alpha$ polarization properties of the (albeit 
small) NLS1 sub-sample are indistinguishable from those of the 
`normal' Seyfert 1s.

\subsubsection{Polarization Variability}
\label{polvar}

We have observed several objects at two or more epochs and as noted 
in Section~\ref{results}, there is evidence in some cases for significant 
polarization variability.  We have discussed the polarization 
variations that occured in Mrk\,509 between 1996 August and 1997 May 
in a previous paper (Young et al.  1999).  In this case, there was a 
decrease in the degree of polarization, with a corresponding decline 
in the very broad H$\alpha$ component present in the polarized flux 
spectrum.  There is also evidence for a small swing in polarization 
PA. The relatively short timescale over which these changes took 
place implies the presence of a compact scattering region.

Of the other objects, the most convincing case for a real change in 
polarization can be made for Akn\,120 (Section~\ref{a120}).  Both 
observations of this source, in 1997 February and 1998 October, were 
obtained at the WHT with almost exactly the same instrumental set-up 
(a slightly larger slit width was used for the later observation).  In 
the interval between the two observations, there was an overall 
decrease in the degree of continuum polarization, accompanied by 
significant changes in the polarization structure over the broad 
H$\alpha$ line, notably the disappearance of the polarization minimum 
associated with the line core and the appearance of a polarization 
peak in the far blue wing.  The continuum polarization appears to have 
dropped to the level of the line core, leaving behind the more highly 
polarized feature in the blue wing.  There does not appear to have 
been any significant change in the polarization PA between the two 
epochs.  A later (1999 August) observation obtained by Schmid et al.  
(2000) shows polarization structure similar to that in our 1997 
February spectrum, suggesting that the large changes that occurred in 
the interval between our two observations have been reversed.  These 
dramatic changes provide strong confirmation of earlier reports (M96) 
that Akn\,120 undergoes polarization variations.

Another object in which a change in polarization seems to have 
occurred is NGC\,7469. The average levels of polarization in this 
object are consistent with what we would expect for interstellar 
polarization and at 2 out of the 3 epochs at which we observed this 
source, the polarization spectra are featureless, also as would be 
expected if the polarization is largely interstellar in origin. 
However, the data from the third epoch (1998 October) reveal a 
significant local peak in percentage polarization associated with the 
red wing of the broad H$\alpha$ line. The presence of this peak is 
presumably the reason why the average H$\alpha$ polarization is a 
factor $\sim 2$ higher than in the 1997 June data, which were 
obtained with an identical set-up (using the WHT) and are of 
comparable quality. The AAT data obtained in 1997 August are of lower 
resolution and quality, but are generally consistent with those of 
1997 June. The polarization peak cannot be attributed to any obvious 
instrumental or spurious observational effects (e.g., cosmic rays, 
CCD defects) and it is difficult, therefore, to avoid the conclusion 
that it represents a real variation in the source. 

In Section~\ref{results} we also compared our polarization 
measurements with those of other authors.  In most cases, the various 
sets of measurements are reasonably consistent.  The only object for 
which we found significant differences is NGC\,3516.  Our 
observations, taken in 1997 February, yield a low level of 
polarization (0.15 per cent), consistent with the expected 
interstellar polarization, whereas in 1994 April, M96 measured 0.5 per 
cent polarization at a PA 40$\degr$ different from ours.  It is 
possible that this discrepancy represents a real variation, the 
intrinsic polarization having decreased, by the time of our 
observation, to the point where the interstellar polarization is 
dominant.  Alternatively, the differences could arise from a 
difference in slit width and orientation.  M96 used a wider slit (2.4 
arcsec) in a different PA and since NGC\,3516 has a bright, 
well-resolved, extended emission line region, this could have resulted 
in contamination by polarized light from an extended scattering medium 
associated with this gas.

Nevertheless, the observations of Mkn\,509, Akn\,120 and NGC\,7469, 
at least, are sufficiently robust as to provide convincing evidence 
that, in some Seyfert 1 nuclei, the broad H$\alpha$ polarization can 
vary on a timescale $\sim 1$ year. Since the emission line polarization 
can only be due to scattering, this implies that at 
least part of the scattering medium is compact, and probably close to 
the BLR. The variations in percentage polarization could arise from a 
physical change in the scattering medium (e.g., an increase in the 
density of scattering particles) or a change in the line flux 
incident on the scatterers. It is worth noting, in this context, that 
all three objects are well known to undergo significant 
broad-emission line variability, resulting from the light-echo 
delayed response of the BLR to changes in the central ionizing 
continuum luminosity (e.g., Peterson 1999 and references therein). 

\subsubsection{Nuclear luminosity}

Our sample covers a fairly wide range in luminosity; $-16.8 \le M_B \le -27.1$ 
(V\'{e}ron \& V\'{e}ron 1998). Nevertheless, there is no clear evidence 
that the polarization properties are strongly related to luminosity. Even 
though several of the objects in our sample are luminous enough ($M_B 
< -23$) to be classified as quasars, their polarization properties do not 
differ significantly from those of the rest of the sample. It may, 
however, be significant that the group identified as having polar 
scattering characteristics similar to those of Type 2 Seyfert nuclei in 
which polarized broad-lines are detected (Fairall 51, NGC4593 and Was 45) 
are among the fainter objects in our sample. Respectively, these objects have 
magnitudes $-$19.8, $-$19.7 \& $-$21.0, compared with an average $M_B \approx -22.6$ 
for the objects listed in Table~\ref{tab:rotsig}, which have polarization properties 
consistent with equatorial scattering. The slightly lower luminosities 
of the former group are consistent with our picture in which the line-of-sight 
to the nucleus is subject to a moderate amount of extinction (Sections~\ref{polstr} 
and~\ref{eq+pol}).

\subsubsection{Host galaxy inclination}

It is also of interest to consider polarization properties in relation to
the inclination of the host galaxy.  There is no discernable relationship
between the principal axis of the active nucleus, as defined by the radio
jet, and the dynamical axis of the parent galaxy among Seyferts (e.g.,
Kinney et al. 2000).  However, the inclination of the host is likely to
govern the amount of obscuring material in the galactic disk, that 
is present along the line-of-sight to the nucleus and this may have some 
bearing on our analysis. Indeed, the apparent lack of Seyfert 1 nuclei in 
edge-on host galaxies suggests that dust within the disk can have a similar 
effect to the torus, obscuring the direct view of the nucleus (e.g., Keel 1980; 
Schmitt et al.  2001).  Among our sample, however, there is no evidence that the 
inclination of the disk of the host galaxy (as measured by its aspect ratio) 
affects the nuclear polarization properties. For example, Mrk\,6 and Fairall 51,
are probably the most clear-cut cases of, respectively, equatorial and
polar scattering. As such, their nuclei are presumably viewed at rather
different orientations. However, their host galaxies have similar aspect
ratios, implying similar inclinations.

\subsection{Inclination-dependent scattering in Seyfert 1 nuclei}
\label{model}

Broadly speaking, the intrinsically polarized objects in our sample 
fall into two groups.  A minority (3 out of 20) exhibit polarization 
properties very similar to those of Type 2 Seyferts (polarization 
increases to shorter wavelengths, with local increases over the broad 
Balmer lines, the polarization PA is constant over the lines), which 
can be attributed to far-field polar scattering.  Most of the 
remainder are characterized by decreases in polarization within the 
core of the broad H$\alpha$ line, in some cases flanked by increases 
in one or both wings.  Many of this group also exhibit significant 
blue-to-red PA rotations across the line profile.  It is clear from the discussion in 
Sections~\ref{parot} and~\ref{polstr}, that both of these 
characteristics (line core depolarization and PA rotation) can be 
produced by scattering of broad-line emission from a rotating disk, if 
the scattering medium lies in the plane of the disk and closely 
surrounds it.  Here we outline a model based on this idea in more 
detail.  Detailed calculations of the polarization produced by this 
scattering configuration will be presented in a forthcoming paper 
(Smith et al., in preparation).

\subsubsection{Equatorial scattering model}
\label{eqscat}

\begin{figure*}

\hspace*{20pt}

\psfig{figure=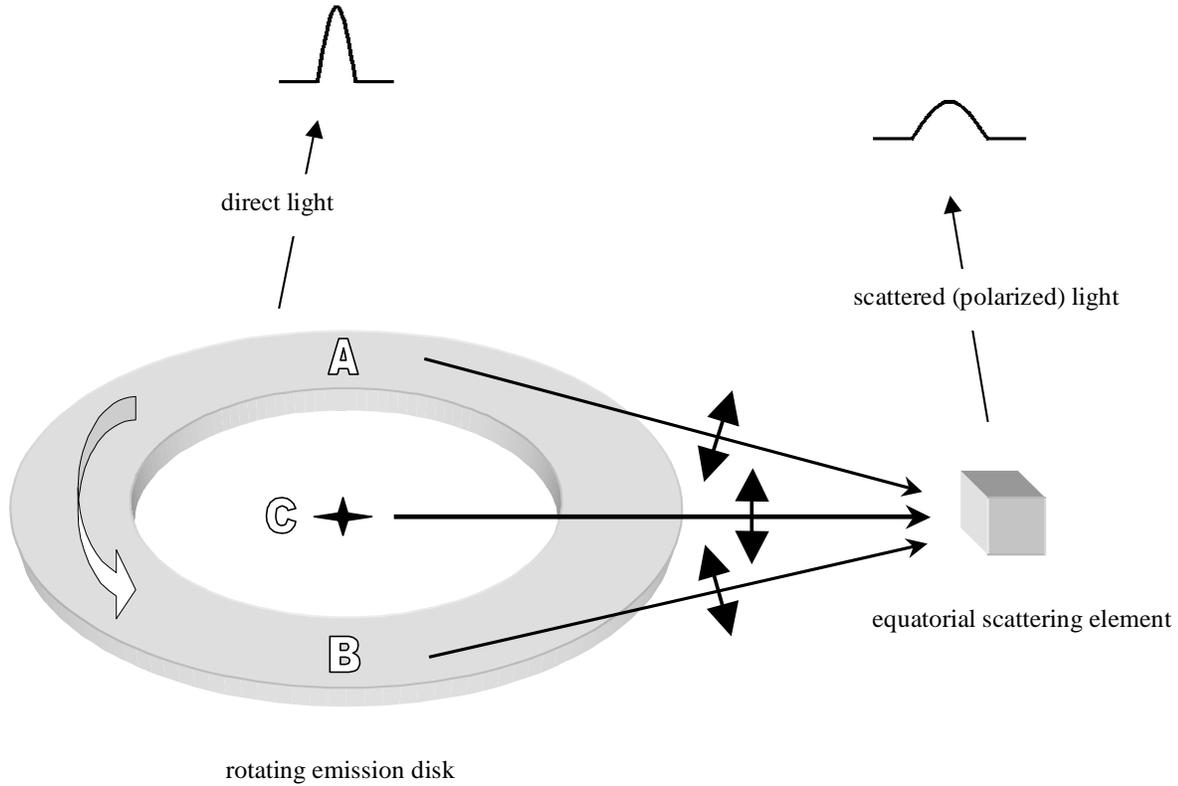,width=6.5in}

\centering{\caption{Schematic illustration of the equatorial 
scattering model.  The disk rotation axis is inclined at some 
intermediate angle to the observer's line-of-sight (which is 
perpendicular to the page). The single scattering element shown is 
part of a co-planar ring surrounding the disk. The thick arrows 
represent the polarization {\bf E} vectors, after scattering, of rays 
originating from, respectively, opposite sides, A and B, of the 
emission disk and a point-like central continuum source, C. The line 
profiles in direct 
and scattered light are also shown.\label{fig:eqscat}}}

\end{figure*}

The main features of the model are sketched in Fig.~\ref{fig:eqscat}.  The 
line-emitting disk is centred on a point-like continuum source, C, and 
is tilted at an arbitrary inclination to the line-of-sight.  A 
co-planar ring of scatterers closely surrounds the disk.  Both the 
disk and ring are surrounded by, and lie in the equatorial plane of, 
the dusty molecular torus that is the key element of Seyfert 
unification schemes.  The disk, ring and torus are, therefore, 
co-axial and we assume that this `system' axis also defines the radio 
axis.

For simplicity, we first consider a single scattering element aligned 
with the major axis of the disk seen in projection.  Light from the 
disk and from the central continuum source is scattered from this 
point in the ring into the observer's line-of-sight and thereby 
becomes linearly polarized.  The {\bf E} vector of the scattered 
continuum light will be parallel, in projection, to the system axis 
(and hence the radio axis).  The scattering element will `see' blue- 
and red-shifted emission from points A and B, respectively, while the 
corresponding {\bf E} vectors are directed either side of the 
continuum {\bf E} vector with equal but opposite offset angles.  The 
result is a rotation in polarization PA from blue to red across the 
line profile, centred on the continuum PA.

In general, the blue-to-red PA rotation is still present even when 
scattering from the complete ring is considered.  When the disk is 
viewed face-on, circular symmetry ensures complete cancellation of the 
polarization produced by any scattering element by that of its 
orthogonal counterpart.  However, when the disk is inclined, light 
scattered from elements aligned with the minor axis of the projected 
disk (opposite points A and B) is less completely polarized, due to 
the smaller scattering angles, than light scattered by elements 
aligned with the major axis.  This breaks the symmetry and leaves a 
net polarization with a PA rotation similar to that expected for a 
single scattering element aligned with the major axis.\footnote{The 
scattering element at the opposite `pole' of the major axis polarizes 
blue- and red-shifted emission at the same PA's as its counterpart, 
since the Doppler shifts for A and B are reversed in sign.}

The model also predicts a variation in the degree of polarization with 
velocity shift across the line profile.  A simple rotating disk 
produces a line profile whose width varies as $\sin{i}$, where $i$ is 
the inclination of rotation axis to the line-of-sight.  Since the 
scattering ring has an edge-on view of the disk ($i=90\degr$), the 
scattered line profile encompasses the whole range in rotational 
velocity present in the disk.  This will not be the case for the 
direct line-of-sight to the observer, which for a Seyfert 1 nucleus, 
must pass within the torus opening.  Observational evidence suggests 
that the average torus opening angle in Seyfert galaxies is $\sim 
45\degr$ (e.g., Lawrence 1991), implying that in a Type 1 Seyfert 
the direct line-of-sight must have $i\la 45\degr$.  It follows that 
the line profile in direct, unpolarized, light has a width 
$\la 1/\sqrt{2}\times$ that in scattered, polarized, light.  The 
combination of scattered and direct line emission will, therefore, 
result in wavelength-dependent dilution of the polarized component.  
With the further addition of the underlying continuum, whose 
polarization is wavelength independent, the observed percentage 
polarization will increase relative to the continuum level in the line 
wings, but will drop to a minimum below that level in the core.

To summarize, we expect that the polarized light produced by this 
scattering configuration will exhibit the following general features 
to some degree:

\begin{enumerate}

\item the position angle of polarization will rotate from blue to 
red across the broad-line profile (provided that the disk subtends a 
finite angle at the scattering ring);

\item the underlying continuum will be polarized at a PA intermediate 
between those of the blue and red line wings. The continuum PA will 
be aligned with the projected disk rotation axis and hence also the 
radio axis;

\item the degree of polarization peaks in the line wings and passes 
through a minimum in the line core.  The underlying continuum is 
polarized at an intermediate level.

\end{enumerate}

As we have already noted, characteristics similar to those summarized 
above are observed in several of our objects.  Table~\ref{tab:rotsig} lists ten 
galaxies whose broad H$\alpha$ lines show either a clear blue-to-red 
rotation in polarization PA across the line profile (e.g., 
Fig.~\ref{fig:a120cs}) 
{\em or} a decrease in the degree of polarization at the line core, 
accompanied by a peak in polarization, above the continuum level, in 
either line wing (e.g., Fig.~\ref{fig:m6}).

\begin{table*}

\centering


\caption{Objects exhibiting polarization characteristics in the broad 
H$\alpha$ line consistent with the equatorial scattering model.}

\begin{tabular}{@{}rccc@{}}

\hline

Object			&PA rotation & Depolarization in   	& 
Polarization peak(s)\\

			&across line profile ?& line core ?	& in either line wing ?	\\

\hline

Akn 120	Epoch 1		& YES		& YES 				& NO 	\\ 

	Epoch 2		& YES 		& NO  				& YES	\\

IZw1    Epoch 1		& NO		& YES				& NO	\\

	Epoch 2		& YES		& YES				& NO	\\

KUV 18217+6419		& YES		& NO				& YES	\\		

Mrk 6			& YES		& YES				& YES	\\

Mrk 304			& NO		& YES				& YES	\\		

Mrk 509	All Epochs	& NO		& YES 				& YES	\\	

Mrk 841			& NO		& YES				& YES	\\

Mrk 876			& YES		& YES				& NO	\\

Mrk 985			& YES		& YES				& NO	\\

NGC 3783		& NO		& YES 				& YES	\\

\hline

\end{tabular}

\label{tab:rotsig}
\end{table*} 

Few of these objects exhibit {\em all} of the expected polarization 
characteristics.  However, the PA rotation and percentage polarization 
structure result from different mechanisms (an anisotropy in 
scattering angle distribution and differential dilution, respectively) 
and, although both effects depend on the line-emitting disk being 
inclined to the line-of-sight, they are also subject to other, 
independent, parameters. Therefore, these 
characteristics would not necessarily be strongly correlated.

Mrk\,6 provides the best overall match to the 
qualitative predictions. The most problematic objects are Akn\,120, 
whose percentage polarization spectrum exhibits very different 
structures at different epochs, and Mrk\,304, in which the H$\alpha$ 
polarization structure agrees quite well with that predicted, whereas 
the polarization PA is nearly orthogonal to the reported radio axis, 
consistent with polar rather than equatorial scattering.

A blue-to-red PA rotation is seen in only 6 out of the 10 
objects.  However, increasing the scattering ring radius relative to 
that of the disk will decrease the angle subtended by the disk at a 
given scattering element and hence reduce the PA swing.  Therefore, 
the absence of a significant PA rotation in some objects may simply 
mean that the scattering ring has a relatively large radius in those 
cases.  When a PA swing {\em is} seen, we would expect the continuum 
polarization PA to bisect the blue-to-red rotation across the line.  The 
continuum PA lies within the range covered by the H$\alpha$ 
polarization PA in 5 out of the 6 objects that exhibit a PA swing 
across the line, but only in two (Mrk\,876 and Mrk\,985) is it 
approximately centred with respect to the blue and red wing PA's.

In only two galaxies, Mrk\,6 and Mrk\,509, do the observed variations 
in the degree of polarization across H$\alpha$ closely match the model 
prediction in that the central dip in polarization is flanked by red 
and blue peaks of similar amplitude.  In general, the prominence of 
the blue- and red-wing polarization peaks varies considerably, as does 
their relative amplitude.  These features are absent in 3 objects 
(I\,Zw1, Mrk\,876, Mrk\,985; which, however, do exhibit PA rotations) 
while, in the others, the blue-wing peak is usually the stronger. 

Mrk 304, whilst showing a peak-trough-peak variation in $p$ over broad H$\alpha$, has an optical polarization PA orthogonal to that of the reported radio axis PA -- apparently consistent with polar rather than equatorial scattering. Nevertheless, it may still be possible to reconcile the polarization characteristics of Mrk 304 with equatorial scattering. Corbett et al. (1998) show that scattering from the inner, far wall of the circum-nuclear torus can produce scattered light whose polarization PA is perpendicular to the radio axis if an object is inclined such that the observer has a clear view of the emission disk and the far inner wall of the torus, but the scattering material at the nearer inner wall and sides of the torus is obscured by the bulk of the torus material. A similar configuration may explain the optical polarization of Mrk 304 in the context of the equatorial scattering model developed in this work. For example, the torus may partially obscure the equatorial scattering region such that most of the scattering material located at the sides and along the nearer segment (i.e., opposite point B in Fig.~\ref{fig:eqscat}) of the scattering ring is obscured, while the emission source and the far side of the scattering ring remain visible. The scattered profile would be broader than in the direct view, so a peak-trough-peak variation in $p$ would be seen across the emission-line. In addition, the scattered flux from along the far side of the equatorial scattering ring (i.e., opposite point A in Fig.~\ref{fig:eqscat}) will have an average PA that is perpendicular to the emission disk axis and hence radio axis. A similar explanation may apply to Akn 564 and NGC 5548, however, the lack of polarization peaks in the line wings remains a problem for these objects. 

Clearly, the polarization features predicted by the equatorial 
scattering model are present to very differing degrees in different 
objects.  This is, perhaps, not surprising given the simple geometry 
of the model and the likely complexity of the emission and scattering 
regions in reality.  For example, the fact that the broad H$\alpha$ 
lines commonly have asymmetric profiles shows that the emission region 
is more complex than the simple disk envisaged in the model.  
Additional emission components or scattering centres, perhaps with 
velocity shifts, will undoubtedly distort the polarization signature 
of the disk/ring system.  For example, a scattering outflow along the 
system axis would produce red-shifted scattered light, polarized 
orthogonal to the equatorial component.  We might thus expect the 
amplitude of the red-wing polarization peak to be reduced (by 
cancellation) relative to that of the blue-wing and its PA of 
polarization to be preferentially modified, breaking the symmetry of 
the blue- and red-wing polarization PA's relative to that of the 
continuum.  Such an outflow might be identified with the polar 
scattering region which, as we discuss below, must also be present in 
Seyfert 1 galaxies.

\subsubsection{Combined equatorial and polar scattering}
\label{eq+pol}

While the equatorial scattering model can account for the most 
striking features that are present in our data, namely the blue-to-red 
polarization PA rotation and the peak-trough-peak variation in degree 
of polarization over the broad H$\alpha$ line, there is also 
considerable evidence that polar scattering is important in Seyfert 1 
nuclei.  In particular, 3 objects have polarization PA's perpendicular 
to the radio source axis and 3 more exhibit polarization spectra very 
similar to those observed in `polarized broad-line' Seyfert 2 nuclei.  
The latter objects, along with others previously described (GM94), 
represent direct observational evidence that the `scattering mirror', 
which is located above the circum-nuclear torus and allows us to 
detect the obscured BLR in Seyfert 2 galaxies, is also present in Type 
1 nuclei.  This is, of course, also implied by the Seyfert unification 
scheme, according to which, similar scattering regions should be 
present in all Seyfert nuclei.

Consideration of the optical polarization properties of Seyfert 1 nuclei
should, therefore, include far-field polar scattering as well as the
near-field equatorial component.  Our detailed model of Mrk\,509 (Young 
2000) includes both of these components and also takes into 
account several other effects, such as scattering off the inner wall of the torus.
Here, for simplicity, we discuss only the combination of near-field
equatorial scattering, as described in Section~\ref{eqscat}, and polar scattering. 
We envisage the polar scattering region as a cone, containing dust or free
electrons, which is co-axial with the torus and has a comparable opening
angle.  

The observed polarization produced by the combination of equatorial and
polar scattering will, in general, depend on the inclination of the system
axis to the line-of-sight.  At very small inclinations (with the system
axis closely aligned with the line-of-sight), we expect a low degree of
polarization, since the circular symmetry ensures that both components of
scattered light will suffer almost complete cancellation.  As the
inclination is increased from $i=0\degr$ to $i=90\degr$, the degree of
polarization increases monotonically for both components.  Calculations
using Young's (2000) scattering code show that, for the wide scattering
cone envisaged, $p$ increases more rapidly with inclination for the
equatorial than for the polar component.  Without the optically thick
torus, therefore, the equatorial component would dominate the observed
polarization at all inclinations.  However, for inclinations $i \ga
45\degr$, the line-of-sight to the BLR and the equatorial scattering
region is blocked by the torus.  We would then observe a Seyfert 2 nucleus,
with broad Balmer lines visible in polarized light as a result of
scattering in the polar region.  In order for polar scattering to make a
sigificant contribution to the polarization in {\em Type 1} Seyferts, the
equatorial component must be suppressed relative to the polar component. 
An obvious possibility is that, in the objects that exhibit signatures of
polar scattering, the direct line-of-sight to the AGN passes through the
(presumably relatively tenuous) upper layers of the torus.  Light from the
the equatorial scattering region might easily suffer sufficient extinction
to allows polar scattering to dominate the observed polarization.  Direct
light from the AGN would, of course, also suffer extinction and as we have
already noted in Section~\ref{polstr}, this would help to explain the polarization
wavelength dependence in objects that have `Seyfert 2-like' polarization
properties (Fairall\,51, NGC\,4593 and, to a lesser extent, Was\,45).

The above analysis suggests that the wide range in polarization properties
displayed by the sample as a whole can be at least partly understood in
terms of an orientation sequence.  Some of the objects that are evidently
dominated by interstellar polarization have low measured polarizations
($\le 0.3$ per cent) and must, therefore, be intrinsically weakly
polarized.  In our scheme, these objects (e.g., Mrk\,926, PG\,1211+143)
would have near pole-on orientations.  The objects listed in 
Table~\ref{tab:rotsig}, which
exhibit H$\alpha$ polarization features characteristic of equatorial
scattering, have intermediate inclinations to the line-of-sight.  Objects
that show evidence of polar scattering (polarization PA perpendicular to
radio axis, or Seyfert 2-like percentage polarization spectra) are viewed
at larger inclination angles, comparable to the torus opening angle, and
suffer a moderate amount of extinction in the upper layers of the torus. 
With a further increase in inclination, the line-of-sight to the nucleus
passes through the main body of the dusty molecular torus and a `polarized
broad-line' Seyfert 2 galaxy is observed.
 
It is worth briefly considering NLS1's in relation to the scheme 
outlined above.  If the small widths of their Balmer lines are due to 
the line-emitting disk having a face-on orientation, the combined 
scattering model implies that they should be weakly polarized.  In 
fact, in our sample, only two (PG\,1211+143 and Mkn\,896) have 
measured polarizations $<0.3$ per cent.  Of the others, 3 are probably 
dominated by interstellar polarization (at levels $\sim 0.5$ per cent) 
and the remaining 3 are intrinsically polarized at similar levels to 
those of the `normal' Seyfert 1s.  One of the latter, Akn\,564, is 
polarized perpendicular to the radio axis, indicating that polar 
scattering is important and hence that the object is viewed at a 
fairly large inclination.  The NLS1's, therefore, appear to be 
distributed amongst  all three of the postulated orientation 
classes, rather than being confined to the pole-on group.

While the combination of equatorial and polar scattering outlined 
above can broadly explain the range of H$\alpha$ polarization 
properties found in the sample, there are a number of intrinsically 
polarized objects whose status within this scheme is unclear.  Most 
of these show a dip in polarization over the broad H$\alpha$ line core but 
there is little evidence for any of the other features associated with 
the emission disk/scattering ring model.  Better data might, perhaps, 
reveal such features in some cases (e.g., MS\,1849.2$-$7832, Mrk\,279 
and Mrk\,290) but in other objects (e.g., NGC\,5548 and Mrk\,335), 
there are large variations in the degree and PA of polarization over 
H$\alpha$ line profile that are difficult to understand in terms of 
the simple scenario presented here.

\section{Conclusion}
\label{conclusion}

We have detected polarization intrinsic to the active nucleus in 20 
out of 36 Seyfert 1 galaxies.  We have measured a significant level of 
polarization in a further 7 objects but in these we cannot exclude the 
possibility of serious contamination by line-of-sight interstellar 
polarization.  Overall, the sample exhibits a wide diversity in 
polarization properties.  The main characteristics of the 
intrinsically polarized objects may be summarized as follows.

\begin{enumerate}

\item Large variations in both the degree and PA of polarization with 
line-of-sight velocity across the broad H$\alpha$ emission line are 
common.

\item 15 objects exhibit a change in polarization PA over the broad 
H$\alpha$ line with approximately half of these showing PA rotations 
from blue-to-red within the H$\alpha$ profile.

\item The broad H$\alpha$ line is usually depolarized relative to the 
adjacent continuum but in many cases the dip in polarization is 
associated mainly with the core of the line profile and is often 
flanked by polarization peaks in the red and blue wings, the blue wing 
peak usually being the most prominent.

\item In 4 objects there is an increase in percentage polarization 
over the broad H$\alpha$ line relative to the adjacent continuum, and 
in the two for which we have sufficient wavelength coverage, the 
continuum polarization rises to the blue.

\item In general, there appears to be no simple relationship 
between the velocity dependences of the polarization PA and the 
degree of polarization across the H$\alpha$ profile, in the sense 
that the presence of a blue-to-red PA rotation does not uniquely 
correspond to any particular type of percentage polarization 
structure.

\item Several objects were observed at two or more epochs and in 3 of 
these (Mrk\,509, Akn\,120 and NGC\,7469) there is convincing evidence 
that the polarization of the broad H$\alpha$ line varies over a 
timescale of $\leq 1$ year. 

\item Of the 10 objects for which radio PA's are available, the
average position angles of polarization are approximately parallel to 
the radio axis in 6 objects and approximately perpendicular in the 
remaining 3.

\item In a number of the `parallel' objects, the blue-to-red rotation in 
polarization PA across the H$\alpha$ profile crosses the radio 
source PA. In some extreme cases (Akn\,120 and KUV\,18217+6419) the 
polarization {\bf E} vector changes alignment across the profile from 
parallel to perpendicular to the radio axis.

\end{enumerate}

We argue that much of the observed diversity in polarization 
properties, including the main features of the distinctive 
polarization structures associated with the H$\alpha$ line, can be 
explained by a model in which the broad Balmer line emission comes 
from a rotating disk and is scattered by two separate scattering 
regions producing orthogonally polarized light. These are: 

\begin{enumerate}

\item a near-field scattering ring co-planar with the line-emitting 
disk and situated within the torus, in its equatorial plane;

\item a far-field scattering cone situated outside the torus but 
aligned with 
the torus/disk axis.

\end{enumerate}

The relative importance of these two scattering regions as sources of 
polarized light is governed primarily by the inclination of the 
system axis to the line-of-sight. The polarization properties of Type 
1 Seyfert nuclei can be broadly classified in an orientation 
sequence. Thus, 

\begin{enumerate}
\item weakly polarized objects are near pole-on systems;

\item objects polarized parallel to the radio axis and exhibiting 
signatures of equatorial scattering in H$\alpha$ polarization (i.e., a 
peak-trough-peak variation in percentage polarization and/or a 
blue-to-red PA rotation across the line profile), are viewed at 
intermediate orientations and are dominated by equatorial scattering;

\item objects polarized perpendicular to the radio axis and 
exhibiting an increase in polarization to short wavelengths, with 
local peaks associated with the broad emission-lines, are viewed at 
grazing incidence to the circum-nuclear torus and are dominated by 
polar scattering. 

\end{enumerate}

The latter category, of which we have identified 3 possible members amongst
our sample, with a further example from the literature, have polarization
properties that are essentially identical to those of Seyfert 2 galaxies in
which polarized broad-lines have been detected.  As in Type 2's, the
presence of an underlying, unpolarized continuum is required to explain
both the continuum polarization wavelength dependence and the increases in
degree of polarization over the broad-lines.  The equatorial component of
scattered light must also be suppressed for polar scattering to
dominate the observed polarization.  Both conditions can
be satisfied if the direct line-of-sight to the BLR and equatorial scattering
region passes through the upper layers of the torus, so that the nuclei of
these objects suffer a higher degree of extinction than other Seyfert 1s. 
This group represents direct observational evidence that a Seyfert 2-like
polar scattering region (the `mirror' that allows us to detect the hidden
BLR in some objects) is also present in Seyfert 1 galaxies.

Type 2 Seyfert galaxies can, themselves, be straightforwardly 
incorporated into this scheme, since at yet larger inclinations, the 
direct line-of-sight to both the BLR and the equatorial scattering 
region is blocked by the torus and a Seyfert 2 nucleus with polarized 
(polar-scattered) broad-lines will be observed.

The narrow-line Seyfert 1 galaxies in our sample display a range of 
polarization properties that is not strikingly different from that of 
the sample as a whole. There is no compelling evidence that they are 
viewed from a preferred orientation, or that their broad-line regions 
are partially obscured.

The model we have outlined, while providing an explanation for the 
key features of our data, cannot in its basic form account for the 
entire range of polarization behaviour exhibited by individual 
objects. There are many possible effects that may modify the 
H$\alpha$ polarization from that predicted by the model. 
Nevertheless, it serves as a useful framework for a general 
understanding of the polarization properties of Seyfert 1 galaxies 
and provides an insight into the relationship between the scattering 
geometries of Type 1 and 2 Seyfert nuclei. Detailed scattering 
calculations based on this model will be presented in future papers.

\section*{Acknowledgements}

JES acknowledges financial support from a PPARC studentship. AR 
thanks the Royal Society for financial support. The WHT is operated 
on the island of La Palma by the Isaac Newton Group in the Spanish 
Observatorio del Roque de los Muchachos of the Instituto de 
Astrofisica de Canarias. We thank the staff of the AAT for help in 
performing the observations at this telescope. The work reported in 
this paper was partly carried out using facilities and software 
provided by the Starlink project. This research has made use of the 
NASA/IPAC Extragalactic Database (NED) which is operated by the Jet 
Propulsion Laboratory, California Institute of Technology, under 
contract with the National Aeronautics and Space Administration.

\end{document}